\documentclass[12pt]{article}
\usepackage{epsf}
\usepackage{epsfig}

\def\slash#1{#1 \hskip-0.45em /}

\def\beq{\begin{eqnarray}}
\def\eeq{\end{eqnarray}}

\begin{document}

\begin{titlepage}
\begin{flushright}
 PITHA 00/20\\
 hep-ph/0008255\\
 24 August 2000\\
\end{flushright}
\vskip 2.4cm

\begin{center}
\boldmath
{\Large\bf Symmetry-breaking corrections to heavy-to-light\\[0.3cm] 
$B$ meson form factors at large recoil}
\unboldmath 
\vskip 2.2cm
{\sc M. Beneke}\hspace*{0.3cm}and\hspace*{0.3cm}{\sc Th.~Feldmann}
\vskip .5cm
{\it Institut f\"ur Theoretische Physik~E, RWTH Aachen, 
52056 Aachen, Germany}

\vskip 2.5cm

\end{center}

\begin{abstract}
\noindent 
Recently it has been shown that symmetries emerging in the 
heavy quark and large recoil energy limit impose various relations  
on form factors that parametrise the 
decay of $B$ mesons into light mesons. 
These symmetries are broken by perturbative effects. In this paper 
we discuss the structure of heavy-to-light form factors including 
such effects and compute symmetry-breaking corrections 
to first order in the strong 
coupling. As an application of our results we consider the 
forward-backward asymmetry zero in the rare decay 
$B\to V \ell^+\ell^-$ and the possibility to constrain 
potential new physics contributions to the Wilson coefficient $C_9$.
\end{abstract}

\vskip 2.5cm

\centerline{\it (submitted to Nucl. Phys. B)}

\vfill

\end{titlepage}


\section{Introduction}

The form factors which we discuss in this paper are matrix elements 
of bilinear quark currents between a $B$ meson and a light pseudoscalar 
or vector meson. These form factors encode strong interaction 
effects in exclusive, semi-leptonic or radiative $B$ decays, such as 
$B\to\pi \ell\nu$, $B\to K^* \gamma$ etc. They also appear as 
non-perturbative parameters in the factorization theorem for 
non-leptonic $B$ decays in the heavy quark mass limit 
\cite{Beneke:1999br,Beneke:1999br2}. 
The knowledge of these form factors therefore helps us to determine 
the CKM coupling $|V_{ub}|$, and to predict CP violating asymmetries 
and other quantities in rare $B$ decays. 

Form factors for heavy-light transitions are presumably dominated 
by QCD interactions at small momentum transfer  
and therefore not computable in perturbation theory. Charles {\em et al.}\/ 
have shown that certain symmetries apply to this soft contribution,
when the momentum of the final light meson is large \cite{Charles:1998dr}. 
These symmetries  
reduce the number of independent form factors from ten to three, but 
they are broken by radiative corrections. In this paper we give 
a brief derivation of these {\em large~recoil}\/ 
symmetry relations and then compute the symmetry-breaking corrections
at first order in the strong coupling constant $\alpha_s$. (At small 
recoil the standard heavy quark symmetries apply \cite{Isgur:1989vq,
Isgur:1990ed}. We do not 
discuss this kinematic region in this paper.) 
This can be done since symmetry-breaking 
corrections arise only from short distances. An interesting
application of our result is the forward-backward asymmetry in the 
rare decay $B\to V\ell^+\ell^-$ (where $V$ is a vector meson and $\ell$ a 
lepton). We show that a measurement of the lepton-invariant mass
squared, where this asymmetry vanishes, yields a direct measurement of
the loop induced Wilson coefficient $C_9$ in the 
weak effective Hamiltonian \cite{Buchalla:1996vs}, 
which is almost free of hadronic 
uncertainties even after including $\alpha_s$-corrections. This 
generalises an observation recently made by Ali 
{\em et al.}\/~\cite{Ali:1999mm}.

\section{Derivation and discussion of large-recoil symmetries}

The form factors for $\bar{B}$ decays into a pseudoscalar meson are 
defined by the following Lorentz decompositions of bilinear quark 
current matrix elements:
\begin{equation}
\langle P(p')|\bar q \, \gamma^\mu b |\bar{B}(p)\rangle =
f_+(q^2)\left[p^\mu+p^{\prime\,\mu}-\frac{M^2-m_P^2}{q^2}\,q^\mu\right]
+f_0(q^2)\,\frac{M^2-m_P^2}{q^2}\,q^\mu,
\label{fvector}
\end{equation}
\begin{equation}
\langle P(p')|\bar q \, \sigma^{\mu\nu} q_\nu b|\bar{B}(p) \rangle =
\frac{i f_T(q^2)}{M+m_P}\left[q^2(p^\mu+p^{\prime\,\mu})-
(M^2-m_P^2)\,q^\mu\right],
\label{ftensor}
\end{equation}
where $M$ is the $B$ meson mass, $m_P$ the mass of the pseudoscalar 
meson and $q=p-p'$. The relevant form factors for $B$ decays into vector 
mesons are defined as 
\begin{equation}
\langle V(p',\varepsilon^\ast)| \bar q \gamma^\mu b | \bar{B}(p) \rangle =
 \frac{2iV(q^2)}{M+m_V} \,\epsilon^{\mu\nu\rho\sigma}
 \varepsilon^{\ast}_\nu \, p^{\prime}_\rho p_\sigma,
\label{V}
\end{equation}
\beq
\langle V(p',\varepsilon^\ast)| \bar q \gamma^\mu\gamma_5 b | \bar{B}(p) 
\rangle &=&
  2m_VA_0(q^2)\,\frac{\varepsilon^\ast\cdot q}{q^2}\,q^\mu + 
  (M+m_V)\,A_1(q^2)\left[\varepsilon^{\ast\mu}-
  \frac{\varepsilon^\ast\cdot q}{q^2}\,q^\mu\right]
\nonumber\\[0.0cm]
  &&\hspace*{-2cm}
-\,A_2(q^2)\,\frac{\varepsilon^\ast\cdot q}{M+m_V}
 \left[p^\mu+p^{\prime\,\mu} -\frac{M^2-m_V^2}{q^2}\,q^\mu\right],
\eeq
\vskip0.2cm
\begin{equation}
\langle V(p',\varepsilon^\ast)| \bar q \sigma^{\mu\nu}q_\nu b | \bar{B}(p)
\rangle =
  2\,T_1(q^2)\,\epsilon^{\mu\nu\rho\sigma}\varepsilon^{\ast}_\nu\, 
  p_\rho p^{\prime}_\sigma,
\end{equation}
\begin{eqnarray}
\langle V(p',\varepsilon^\ast)| \bar q \sigma^{\mu\nu} \gamma_5 q_\nu b | 
\bar{B}(p) \rangle&=&
(-i)\,T_2(q^2)\left[(M^2-m_V^2)\,\varepsilon^{\ast\mu}-(\varepsilon^\ast\cdot
q)\,(p^\mu+p^{\prime\,\mu})\right]
\nonumber\\[0.0cm]
 && \hspace*{-2cm}
+\,(-i)\,T_3(q^2)\,(\varepsilon^\ast\cdot
q)\left[q^\mu-\frac{q^2}{M^2-m_V^2}(p^\mu+p^{\prime\,\mu})\right],
\label{ffdef}
\end{eqnarray}
where $m_V$ ($\varepsilon$) is the mass (polarisation vector) 
of the vector meson and we use the sign convention $\epsilon^{0123}=-1$.

We begin with a qualitative discussion of strong interaction symmetries 
and radiative corrections for form factors in $B$ meson decay in the 
limit $M/\Lambda_{\rm QCD}\to\infty$. ($\Lambda_{\rm QCD}$ is the 
strong interaction scale.) It is useful to briefly recapitulate the 
implications of heavy quark symmetry, when the final meson $P(V)$ is 
also heavy, for example a $D$ meson.

\subsection{Recapitulation: heavy-heavy form factors}

As long as the velocity transfer to the $D$ meson remains of order 
1, we may assume that the heavy quarks interact with the 
spectator quark (and other soft degrees of freedom) exclusively via  
soft exchanges characterised by momentum transfers much smaller than 
the heavy quark masses. Any hard interaction would imply large momentum 
of the spectator quark in the $B$ meson or $D$ meson or both, and such 
a configuration is assumed to be highly improbable. The simplifications 
that occur when heavy quarks interact only with soft gluons are 
formalised as heavy quark effective theory (HQET)
\cite{Eichten:1990zv,Grinstein:1990mj,
Georgi:1990um,Neubert:1994mb}. The heavy quark momentum $p_Q$ ($Q=b,c$) 
is decomposed into a 
large ``kinematic'' term and a small residual momentum ($k^\mu$),
\beq
p_Q^\mu = m_Q v^\mu +k^\mu, \qquad  |k| \ll m_Q,
\eeq
where $v$ is the heavy meson velocity. To leading order in 
$\Lambda_{\rm QCD}/m_Q$, the interaction of heavy quarks with soft gluons 
is described by the effective lagrangian
\beq
  {\mathcal L}_{\rm HQET} &=& 
    \bar Q_v \, (i \, v \cdot D) \, Q_v 
      + O(1/m_Q) 
\label{HQET}
\eeq
Here $Q_v(x)=e^{i m_Q v \cdot x}\,\frac{1+\slash v}{2}\, Q(x)$ 
denotes the large components of the heavy quark spinor field with its 
leading $m_Q$-dependence made explicit, and $D^\mu=\partial^\mu - 
i g_s A^\mu$ is the covariant derivative in QCD. 

\begin{figure}[t]
   \vspace{-3.7cm}
   \epsfysize=27cm
   \epsfxsize=18cm
   \centerline{\epsffile{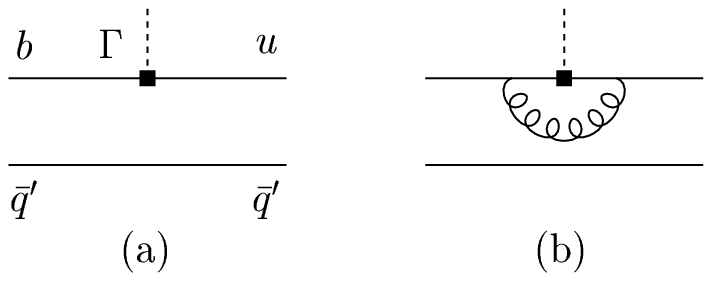}}
   \vspace*{-22.7cm}
   \epsfysize=27cm
   \epsfxsize=18cm
   \centerline{\epsffile{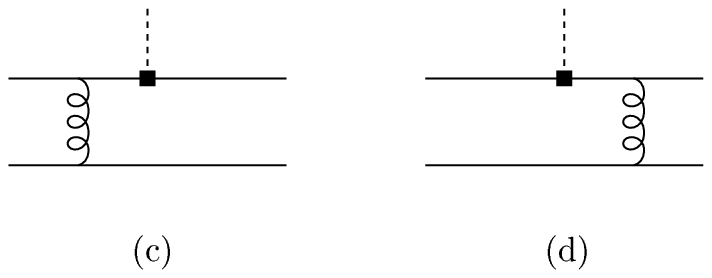}}
   \vspace*{-21.5cm}
\caption[dummy]{\label{fig1}\small Different contributions to the 
$B\to P(V)$ transition. (a) Soft contribution (soft interactions with 
the spectator antiquark $\bar{q}'$ are not drawn). (b) Hard vertex 
renormalisation. (c,d) Hard spectator interaction.}
\end{figure}

Eq.~(\ref{HQET}) implies the well-known spin and heavy flavour symmetries 
which arise in the infinite quark mass limit 
\cite{Isgur:1989vq,Isgur:1990ed}. 
A consequence of these symmetries is that the three pseudoscalar and  
seven vector form factors defined in Eqs.~(\ref{fvector})-(\ref{ffdef}) 
are all related to a single function of velocity transfer $v\cdot v'$, 
$\xi(v\cdot v')$, whose absolute normalization is known at zero recoil 
($\xi(1)=1$) owing to current conservation. The heavy quark symmetries 
are violated by radiative corrections (as well as higher dimension 
operators in the effective lagrangian), such as the one shown in 
Fig.~\ref{fig1}b. (The disconnected spectator quark line in 
Figs.~\ref{fig1}a and \ref{fig1}b is meant to indicate that it is 
connected to the other lines only via soft exchanges. Fig.~\ref{fig1}a 
therefore stands for the leading term in the heavy quark mass limit.) 
The symmetry-breaking effects are caused only by the short-distance 
part of Fig.~\ref{fig1}b. They are accounted for by multiplicatively 
renormalising the heavy quark current in HQET, 
$[\bar{c}\Gamma b]_{\rm QCD} = \sum_{\Gamma'} 
C_{\Gamma\Gamma'}(v\cdot v',\alpha_s) \,
[c_{v'}\Gamma' b_v]_{\rm HQET}$. Hence, neglecting $1/m_Q$ corrections, 
there remain nine parameter-free relations between the pseudoscalar 
and vector form factors. 

The assumption that any interaction with the spectator quark is 
soft means that diagrams such as those in Fig.~\ref{fig1}c and 
\ref{fig1}d must vanish, when some of the gluon's momentum 
components stay finite in the heavy quark mass limit. This is clearly 
an approximation, whose validity depends on the behaviour of the 
$B$ and $D$ meson wave functions. A non-vanishing contribution from 
these diagrams cannot be accounted for by multiplicative renormalisation 
of the heavy quark currents. In fact, it is not accounted for to any order 
in the heavy quark expansion in the HQET formalism. As we shall see, 
a non-vanishing contribution from ``hard spectator interactions'' is 
one of the main differences between heavy-heavy form factors and 
heavy-light form factors at large recoil. 

\subsection{Heavy-light form factors at large recoil}
\label{sec:HLsymm}

We now turn to decays into light mesons and require the light meson 
to have energy $E=(M^2+m^2-q^2)/(2 M)$ 
of order $M/2$. More precisely, we require 
that $E-M/2\ll M$ or, equivalently, $q^2 \ll M^2$. Let us continue 
to assume that the $b$ quark and the energetic $u$ quark, created 
in the $b\to u$ transition, interact with the spectator quark (and 
other soft degrees of freedom) exclusively via soft exchanges. We may 
then continue to use Eq.~(\ref{HQET}) for the $b$ quark. A similar 
simplification occurs for the energetic light quark, for which we may 
use the eikonal approximation. 

We introduce a light-like vector $n_-^\mu$ ($n_-^2=0$)
parallel to the four-momentum $p'$ of the light meson. (In the
following we always neglect effects quadratic in the light meson mass,
so that ${p'}^2\approx 0$. We retain ``kinematic'' corrections 
linear in $m_P$ or $m_V$ introduced through the definition of the 
form factors.) Since we are discussing the soft contribution to the 
form factor, the $u$ quark created in the decay of the $b$ quark 
carries almost all the energy of the light meson, while the spectator 
quark is soft. Hence, we write 
\beq
&&  p_u^{\prime\mu} = E \, n_-^\mu + k'{}^\mu, \qquad |k'| \ll E, 
\eeq
where $p'_u$ is the momentum of the $u$ quark and 
$k'{}^\mu$ is a small residual momentum. To leading order in 
$\Lambda_{\rm QCD}/E$, the interaction of energetic quarks with soft
gluons is described by the eikonal lagrangian 
\cite{Dugan:1991de,Charles:1998dr}
\beq
\label{leet}
  {\mathcal L}_{\rm eik} &=& 
    \bar q_n \, \frac{\slash n_+}{2} \, (i \, n_- \cdot D) \, q_n
   + O(1/E_q),
\eeq
where $q_n(x)=
e^{i E_q n_- \cdot x} \, \frac{\slash{n}_- \slash {n}_+}{4} \,q(x)$
are the large components of the light quark spinor field.  
Here $n_+=2 v-n_-$ is another light-like vector with $n_+ \cdot n_- =
2$ and $E_q\approx E$ is the energy of the light quark. (The factor 
$\slash n_+/2$ in Eq.~(\ref{leet}) can be omitted in the calculation 
of on-shell correlation functions.)

In the large recoil limit, the combination of Eqs.~(\ref{HQET}) and 
(\ref{leet}) implies non-trivial relations between the {\em soft}\/ 
contributions to the form factors \cite{Charles:1998dr}, which we
rederive below. There is an important distinction between the
effective lagrangian for heavy quarks and for energetic light 
quarks. The effective lagrangian (\ref{leet}) applies to light 
mesons produced in an asymmetric configuration, in which a single 
quark carries almost all momentum. Even for light-cone dominated 
processes this is an atypical configuration (the preferred one having 
nearly equal momentum of the quark and antiquark), not speaking of 
the wave functions that diagonalise the strong interaction 
hamiltonian. For this reason, although the interaction (\ref{leet}) 
is spin-symmetric, the symmetry is not realised in the hadronic 
spectrum, and there exists no relation between the soft contributions 
to the form factors of pseudoscalar and vector mesons. Furthermore, 
the probability that such an asymmetric parton configuration
hadronises into a light meson depends on the energy of the meson. 
Hence, the soft contributions to the form factors are energy-dependent
functions, whose absolute normalization is not known. This is to be 
contrasted to the case of heavy-heavy form factors, for which 
spin symmetry relates pseudoscalar and vector mesons, and for which 
the Isgur-Wise form factor $\xi(v\cdot v')$ is independent on the 
heavy quark mass.

To work out the large-recoil symmetry constraints 
on the soft form factor, we 
use a technique familiar from HQET.
Continuing to neglect hard 
interactions, we write $[\bar{q}\Gamma b]_{\rm QCD} = [\bar{q}_n 
\Gamma b_v]_{\rm eff}$. The form factors at large recoil are then 
represented by 
\beq
  \langle L(En_-)| \bar{q}_n \Gamma \, b_v| B(Mv)\rangle
  &=& {\rm tr} \left[A_L(E)\, \overline {\mathcal M}_{\rm L} 
    \, \Gamma \, {\mathcal M}_{\rm B} \right],
\label{trace-form} 
\eeq
where $L=P,V$ and 
\beq
  \overline{\mathcal M}_{\rm L} &=&
            \Bigg\{ \begin{array}{c}
        (-\gamma_5) \\ \slash \varepsilon^* 
         \end{array}\Bigg\}\, \,\frac{\slash n_+ \slash n_-}{4}
        \begin{array}{cl}
        \qquad &\mbox{$L=P$} \\
        & \mbox{$L=V$}
         \end{array} 
\qquad\quad
{\mathcal M}_{\rm B} \,=\,  \frac{1+\slash v}{2} \,(-\gamma_5) 
\label{IW-pi}
\eeq
with $\varepsilon$ the polarisation vector of the vector meson. 
(The generalisation of ${\mathcal M}_{\rm B}$ to the case of a 
vector $B$ meson is obvious. However, in this paper we restrict 
ourselves to pseudoscalar $B$ mesons. In Eq.~(\ref{trace-form}) we assume 
the conventional relativistic normalisation of the states. The factor 
$\sqrt{M}$ which would then normally appear in ${\mathcal M}_{\rm B}$ 
is absorbed into $A_L(E)$.) The function $A_L(E)$ contains the 
long-distance dynamics, but it is independent on the Dirac structure 
$\Gamma$ of the current, because the effective lagrangians
(\ref{HQET}) and (\ref{leet}) do not contain a Dirac matrix. The most
general form $A_L(E)$ can take is therefore 
\begin{equation}
A_L(E) = a_{1L}(E) +a_{2L}(E)\,\slash v+a_{3L}(E)\,\slash n_-+
a_{4L}(E)\,\slash n_-\slash v,
\end{equation}
but the projectors $\overline{\mathcal M}_{\rm L}$, 
${\mathcal M}_{\rm B}$ imply that not all the $a_{iL}(E)$ are 
independent. Accounting for these projectors, the most general 
form is 
\begin{eqnarray}
A_P(E) &=& 2 E \,\xi_P(E), 
\\
A_V(E) &=& E \,\slash n_-\left(\xi_\perp(E)-\frac{\slash
v}{2} \,\xi_\parallel(E)\right)
\end{eqnarray}
with a conveniently chosen overall normalisation. It follows that 
the three pseudoscalar meson form factors are all related to a single
function $\xi_P(E)$ and the seven vector meson form factors are all 
related to two unknown functions, $\xi_\perp(E)$ and
$\xi_\parallel(E)$. The latter two functions are chosen such that 
only $\xi_\perp(E)$ contributes the form factors for a transversely 
polarised vector meson and only $\xi_\parallel(E)$ contributes the 
production of a longitudinally polarised vector meson. Performing 
the trace in Eq.~(\ref{trace-form}), we obtain 
\begin{eqnarray}
&&\hspace*{-0.3cm}
\langle P(p')|\bar q \, \gamma^\mu b |\bar{B}(p)\rangle =
2 E \,\xi_P(E)\,n_-^\mu,
\label{fvector1}
\\
&&\hspace*{-0.3cm}
\langle P(p')|\bar q \, \sigma^{\mu\nu} q_\nu b|\bar{B}(p) \rangle =
2i E\,\xi_P(E)\,\left((M-E) \,n_-^\mu-M v^\mu\right)
\label{ftensor1}
\end{eqnarray}
for pseudoscalar mesons, and 
\begin{eqnarray}
&&\hspace*{-0.3cm}
\langle V(p',\varepsilon^\ast)| \bar q \gamma^\mu b | \bar{B}(p) \rangle =
 2 i E \,\xi_\perp(E)\,\epsilon^{\mu\nu\rho\sigma}
 \varepsilon^{\ast}_\nu \,n_{-\rho} v_\sigma,
\\
&&\hspace*{-0.3cm}
\langle V(p',\varepsilon^\ast)| \bar q \gamma^\mu\gamma_5 b | \bar{B}(p) 
\rangle = 2 E\,\left(\xi_\perp(E)\,(\varepsilon^{\ast\mu}-
\varepsilon^\ast\cdot v\,n_-^\mu)+\xi_\parallel(E)\,
\varepsilon^\ast\cdot v\,n_-^\mu\right),
\label{parperp1}\\
&&\hspace*{-0.3cm}
\langle V(p',\varepsilon^\ast)| \bar q \sigma^{\mu\nu}q_\nu b | \bar{B}(p)
\rangle = 2 E M \,\xi_\perp(E)\,
\epsilon^{\mu\nu\rho\sigma}\varepsilon^{\ast}_\nu\, 
  v_\rho n_{-\sigma},
\\
&&\hspace*{-0.3cm}
\langle V(p',\varepsilon^\ast)| \bar q \sigma^{\mu\nu} \gamma_5 q_\nu b | 
\bar{B}(p) \rangle =
 (-2 i E)\,\Big[\xi_\perp(E)\,M\,(\varepsilon^{\ast\mu}-
\varepsilon^\ast\cdot
v\,n_-^\mu)
\nonumber\\
&&
\hspace*{2cm}+\,\xi_\parallel(E)\,\varepsilon^\ast\cdot v
\,\left((M-E) \,n_-^\mu-M v^\mu\right)\Big]
\label{ffdef1}
\end{eqnarray}
for vector mesons, in agreement with Ref.~\cite{Charles:1998dr}. 
(Note, however, the different convention for the longitudinal 
form factor with $\xi_\parallel(E)=m_V/M\,\zeta_\parallel(E)$, 
$\zeta_\parallel(E)$ being defined in Ref.~\cite{Charles:1998dr}. 
We also neglect some $m_V^2/M^2$ terms that should not be kept at leading 
order in $1/M$.) Comparing 
Eqs.~(\ref{fvector})-(\ref{ffdef}) with 
Eqs.~(\ref{fvector1})-(\ref{ffdef1}), we find the following form
factor relations:
\beq
f_+(q^2) = \frac{M}{2E} \, f_0(q^2) = \frac{M}{M+m_P} \, f_T(q^2) &=& 
\xi_P(E)
\label{pirelation}
\eeq
for pseudoscalar mesons and
\begin{eqnarray}
&&\hspace*{-1.5cm}\frac{M}{M+m_V} \, V(q^2) = \frac{M+m_V}{2 E} \, A_1(q^2) 
       = T_1(q^2) =
       \frac{M}{2E} \, T_2(q^2) = \xi_\perp(E), 
\label{rhorelation1}
\\
\hspace*{-0.5cm}\frac{m_V}{E} \, A_0(q^2) &=&  
\frac{M+m_V}{2 E}\,A_1(q^2) - \frac{M-m_V}{M} \,  A_2(q^2) =  
\frac{M}{2 E}\, T_2(q^2) - T_3(q^2) = \xi_\parallel(E) 
\label{rhorelation2}
\\[-0.6cm]
&&\nonumber
\end{eqnarray}
for vector mesons. These relations are valid for the {\em soft 
contribution}\/ to the form factors at large recoil, neglecting
corrections of order $1/m_b$ and $\alpha_s$. (The $\alpha_s$
corrections will be computed below.) 

We must now examine more carefully our assumption that the $b$ quark
and the $u$ quark created at the weak interaction vertex interact
with the spectator quark exclusively via the exchange of soft gluons. 
As in the case of a heavy-to-heavy transition, there will be a vertex 
correction (of the type shown at one loop in Fig.~\ref{fig1}b). The 
hard part of this vertex correction does not respect the symmetry relations, 
but it can be accounted for in perturbation theory by multiplicatively
renormalising the current $[u_n\Gamma b_v]_{\rm eff}$ in the effective
theory just as in the case of a heavy-to-heavy transition. There is 
an additional complication compared to the heavy-heavy case, because 
there is a long-distance sensitive contribution from energetic gluons 
whose momentum is collinear to the momentum of the outgoing $u$
quark. This ``hard-collinear''  
contribution is not naturally part of the soft form
factor, neither is it perturbatively computable. We shall discuss this
in some more detail later, but the structure of the argument does not 
change due to this extra contribution. 

The important new element of the discussion is provided by the hard
spectator interaction shown 
in Figs.~\ref{fig1}c and \ref{fig1}d. In the absence of 
a hard spectator interaction, the light meson is produced in a 
parton configuration, in which the $u$ quark carries all momentum 
of the meson, up to an amount of order $\Lambda_{\rm QCD}$ in the 
$B$ meson rest frame. In contrast, a hard interaction with the 
spectator quark allows the meson to be formed in a preferred
configuration, in which the momentum is distributed nearly equally
between the two quarks. To estimate the relative importance of the two
contributions, we need to know the amplitude for producing a light
meson in an asymmetric configuration. 

We consider first the hard contribution. (A more extensive version of 
the following discussion can be found in Sect.~3.2 of
\cite{Beneke:1999br2}.) Since both quarks that form the light 
meson have momentum of order $M$ by assumption, and the gluon in 
Figs.~\ref{fig1}c and \ref{fig1}d has 
virtuality of order $M\Lambda_{\rm QCD}$, this contribution can be 
computed by means of the hard-scattering approach to exclusive processes 
\cite{Lepage:1980fj,Efremov:1980qk}. We shall do this explicitly 
below; the resulting scaling behaviour for the pseudoscalar meson 
form factors is 
\begin{equation}
\label{hardbpi}
f_{+,0,T;\,\rm hard}(q^2\approx 0) 
\sim \alpha_s(\sqrt{M\Lambda_{\rm QCD}})\,
\left(\frac{\Lambda_{\rm QCD}}{M}\right)^{\!3/2}.
\end{equation}
The scaling law for the soft contribution can be derived in different 
ways, but all of them make use of the endpoint behaviour of the 
pion's light-cone distribution amplitude \cite{Lepage:1980fj}. If we 
{\em assume}\/ that the distribution amplitude vanishes linearly when
the longitudinal momentum fraction of the spectator quark approaches 
zero, as is suggested by its asymptotic form, we find 
\begin{equation}
\label{softbpi}
f_{+,0,T;\,\rm soft}(q^2\approx 0) \sim \xi_P(E\approx M/2) 
\sim \sqrt{\frac{M}{E}}\left(\frac{\Lambda_{\rm QCD}}{E}\right)^{\!3/2}
\sim \left(\frac{\Lambda_{\rm QCD}}{M}\right)^{\!3/2}.
\end{equation}
(This scaling law was first derived in the context of QCD sum rules 
\cite{Chernyak:1990ag}. Recent evaluations of the sum rule including 
radiative corrections \cite{Khodjamirian:1997ub,Bagan:1997bp} also 
agree with Eqs.~(\ref{hardbpi}), (\ref{softbpi}).) The hard and 
soft contributions to Figs.~\ref{fig1}c and \ref{fig1}d are not 
separately well-defined. The hard-scattering kernel has a logarithmic 
endpoint divergence \cite{Szczepaniak:1990dt,Burdman:1992hg}; one 
must introduce a factorisation scale and factorise the endpoint 
divergence into the soft form factor $\xi_P(E)$. (In the context 
of QCD sum rules this point is also discussed in
\cite{Bagan:1997bp}.). We can summarize this discussion by the 
following, tentative, factorization formula for a heavy-light 
form factor at large recoil, and at leading order in $1/M$:
\beq
\label{fff}
f_i(q^2)  &=& C_i \, \xi_P(E)  + 
\Phi_B \otimes T_i \otimes \Phi_P,
\eeq
where $\xi_P(E)$ is the soft part of the form factor, to which the 
symmetries discussed above apply; $T_i$ is a hard-scattering kernel 
(with the endpoint divergence regulated in a certain manner),
convoluted with the light-cone distribution amplitudes of the 
$B$ meson and the light pseudoscalar meson; $C_i=1+O(\alpha_s)$ 
is the hard vertex 
renormalisation (including, at present, the hard-collinear
contribution discussed above). Eq.~(\ref{hardbpi}) implies that 
the hard spectator interaction (Figs.~\ref{fig1}c and \ref{fig1}d) 
is suppressed by one power of $\alpha_s$
relative to the soft contribution (Fig.~\ref{fig1}a). 
Hence the form factor  
relation (\ref{pirelation}) is indeed correct at leading order 
in $1/M$ {\em and}\/ $\alpha_s$. The correction at order $\alpha_s$
can be computed in terms of a hard vertex renormalisation and 
the hard spectator interaction diagrams. This will be done explicitly 
in Sect.~\ref{symbreak}.

The previous discussion applies unmodified to form 
factors of transversely polarised vector mesons. For 
longitudinally polarised vector mesons, we need 
to keep in mind that $\xi_\parallel(E)/
\xi_\perp(E)\sim m_V/E$. Therefore $m_V/E\times A_0$, $A_2$ and $T_3$ 
scale as $(\Lambda_{\rm QCD}/M)^{5/2}$, when one considers 
longitudinally polarised vector mesons. But since 
the longitudinal polarisation vector is enhanced by a factor 
$E/m_V$ in the large-energy limit, this implies that the form factors 
above times the polarisation vectors and the form factors of
transversely polarised vector mesons all follow the same scaling 
laws in the heavy quark limit, and they all obey a factorisation
formula analogous to Eq.~(\ref{fff}). In particular, the two terms of 
Eqs.~(\ref{parperp1}) and (\ref{ffdef1}) containing $\xi_\parallel(E)$ and 
$\xi_\perp(E)$, respectively, both scale as $M^{1/2}
\Lambda_{\rm QCD}^{3/2}$ in the heavy quark/large energy limit. 

Our discussion up to now has ignored the possibility that the 
configuration in which one quark carries almost all momentum, i.e. the
soft contribution to the form factors, may be suppressed by a 
Sudakov form factor. If this were the case, then the term 
$C_i \, \xi_P(E)$ in Eq.~(\ref{fff}) would be subleading compared 
to the hard spectator term $\Phi_B \otimes T_i \otimes \Phi_P$, 
and perhaps it could be ignored entirely. In this situation the 
form factor would be computable completely
in the hard-scattering approach and 
the symmetry relations (\ref{pirelation})-(\ref{rhorelation2}) 
would seem to be irrelevant. This is assumed, for
example, by the treatment of the $B\to \pi$ form factors in 
Refs.~\cite{Akhoury:1994uw,Dahm:1995ne}. Since in reality the energy of the 
outgoing light quark is not particularly large (around $2.5\,$GeV), 
it is equally possible that the 
Sudakov form factor does not suppress the soft contribution 
sufficiently, so that the symmetry relations remain approximately 
valid. As our default power 
counting we shall therefore take the case that the soft contribution 
dominates by one power of $\alpha_s$, as discussed above.

\section{Calculation of symmetry-breaking corrections}
\label{symbreak}

In this section we compute the corrections of order $\alpha_s$ to the 
symmetry relations (\ref{fvector1})-(\ref{rhorelation2}). As discussed
above these are of two distinct origins and we begin with the more 
familiar case of vertex renormalisation (Fig.~\ref{fig1}b), and then 
turn to the hard spectator interaction.

\subsection{Vertex corrections}

The one-loop diagram in Fig.~\ref{fig1}b contains ultraviolet and 
infrared divergences. The UV divergences are treated by dimensional 
regularisation ($d=4-2\epsilon$); the IR divergences are regulated by 
introducing a (small) mass term $\lambda$ for the gluon, and then 
factored into the soft form factor. 
 
Using standard techniques, we obtain for a generic 
Dirac structure $\Gamma$ at the heavy-to-light vertex the 
following result for the one-loop vertex correction:
\beq
\bar{u}(p') \Gamma(p',p) u(p) &=& 
\frac{\alpha_s \, C_F}{4\pi} \,\bar{u}(p') \,\Bigg[ 
 \Bigg\{ - \frac12\,\ln^2 \frac{\lambda^2 m_b^2}{(m_b^2-q^2)^2}
         - 2\,\ln \frac{\lambda^2 m_b^2 }{(m_b^2 - q^2)^2}
\nonumber\\
&&\hspace*{-3cm}     
         - \,2\,{\rm Li}_2\left[\frac{q^2}{m_b^2}\right]
         - 2 \frac{m_b^2}{q^2}\,\ln\left[1 - \frac{q^2}{m_b^2}\right] 
         - 3 - \frac{\pi^2}{2} \Bigg\} \,\Gamma 
\nonumber\\[0.1cm]
&&\hspace*{-3cm} +\, \frac14 \left\{ \frac{1}{\hat\epsilon} + 
    3 - \ln\frac{m_b^2}{\mu^2} - \left( 1 - \frac{m_b^2}{q^2} \right) \, 
    \ln \left[1-\frac{q^2}{m_b^2}\right]\right\} 
  \,\gamma^\alpha\gamma^\beta \, \Gamma \, \gamma_\beta \gamma_\alpha 
\nonumber \\[0.1cm]
&&\hspace*{-3cm} 
        +\,\frac{ q^2 + m_b^2 \, \ln \left[1-\frac{q^2}{m_b^2}\right]}
       { 2 \, q^4} \, \gamma^\alpha \, \slash p\, \Gamma  
             \, \slash p' \, \gamma_\alpha 
+ \frac{q^2 + (m_b^2-q^2) \, \ln \left[1-\frac{q^2}{m_b^2}\right]}
  { 2 \, q^4} \  m_b \, \gamma^\alpha \, \slash p\, \Gamma \,\gamma_\alpha 
\nonumber \\[1em]
&&\hspace*{-3cm}
  -\,\frac{q^2+(m_b^2-2 q^2)\, \ln \left[1-\frac{q^2}{m_b^2}\right]}
  { q^4} \  m_b \,\Gamma  \, \slash p' \Bigg]\,u(p),
\label{vertex}
\eeq
where $\bar u(p')$ and $u(p)$ denote the external Dirac spinors for
the light and heavy quark, respectively, and we have defined 
$1/{\hat\epsilon} \equiv 1/\epsilon -\gamma_E +\ln4\pi$. For a 
given current $\Gamma$, the product 
$\gamma^\alpha\gamma^\beta \, \Gamma \, \gamma_\beta \gamma_\alpha$ 
is evaluated in the naive dimensional regularisation (NDR) scheme 
with anticommuting $\gamma_5$, and the $1/{\hat\epsilon}$ 
pole is then subtracted. This corresponds 
to defining the bilinear quark current matrix elements in the 
$\overline{\rm MS}$/NDR scheme. 

The coefficients $C_i$ in Eq.~(\ref{fff}) would normally be obtained by 
computing the one-loop vertex correction in the HQET/eikonal 
effective theory, using the same infrared regularisation as in the 
full theory calculation above. The one-loop correction to $C_i$ is 
simply the difference between the two calculations and if both 
theories have the same infrared behaviour, $C_i$ must turn out to be 
independent on the infrared regularisation. This is not the case 
here, because the effective theory does not correctly reproduce 
the hard-collinear infrared divergence. However, Eq.~(\ref{vertex}) 
shows that all infrared divergent terms have the same structure as 
the original current $\Gamma$, so that they can simply be absorbed into 
a redefinition of the the functions $\xi_P$, $\xi_\perp$ and 
$\xi_\parallel$, irrespective of their origin. (The same is true for
the quark self-energy contributions; therefore we do not need to 
calculate them explicitly.) In other words, the hard-collinear 
contributions preserve the HQET/large-recoil symmetries and can hence 
be disregarded in the discussion of symmetry-breaking corrections.

We therefore find it convenient to define the factorisation scheme 
(or renormalisation conventions for the ``soft form factors'') by 
imposing the condition that 
\beq
 && f_+ \equiv  \xi_P, 
  \qquad \ 
 V   \equiv  \frac{M+m_V}{M} \, \xi_\perp, 
  \qquad
  A_0 \equiv  \frac{E}{m_V} \, \xi_\parallel, 
\label{constraint}
\eeq
hold exactly to all orders in perturbation theory. (Such a ``physical'' 
scheme is similar to defining the quark parton distribution to be 
the structure function $F_2$ in deep inelastic scattering to all
orders in perturbation theory.) Having fixed the factorisation 
scheme, we insert, for a given current $\Gamma$, Eq.~(\ref{vertex}) 
into Eq.~(\ref{trace-form}) and express the result in terms of the 
uncorrected (tree-level) soft form factors 
$\xi_P^{(0)}$, $\xi_\perp^{(0)}$ and $\xi_\parallel^{(0)}$. The 
relation between these uncorrected form factors and the corrected 
ones $\xi_P$, $\xi_\perp$ and $\xi_\parallel$ is determined by 
Eq.~(\ref{constraint}) and by eliminating 
$\xi_P^{(0)}$, $\xi_\perp^{(0)}$ and $\xi_\parallel^{(0)}$, we finally
express all other form factors in terms of 
$\xi_P$, $\xi_\perp$ and $\xi_\parallel$. The result reads 
\beq
  f_0 &=& \frac{2E}{M} \, \xi_P \, \left(1 +
    \frac{\alpha_s \, C_F}{4\pi} \,
    \left[ 2 - 2 \, L \right] \right)
  + \frac{\alpha_s \, C_F}{4\pi} \, \Delta f_0, 
\label{add1}
 \\[0.5em]
  f_T &=& \frac{M+m_P}{M} \, \xi_P \, \left(1 +
    \frac{\alpha_s \, C_F}{4\pi} \,
    \left[ \ln \frac{m_b^2}{\mu^2} + 2 \, L \right]
  \right)
+\frac{\alpha_s \, C_F}{4\pi} \, \Delta f_T
\label{vertexcorr1}
\eeq
for the remaining form factors of pseudoscalar mesons and
\beq
A_1 &=& \frac{2 E}{M+m_V} \, \xi_\perp +\frac{\alpha_s \, C_F}{4\pi}
\, \Delta A_1, 
\label{add2} \\[0.5em]
A_2 &=& \frac{M}{M-m_V} \, \left( \xi_\perp - \xi_\parallel \,
  \left( 1 +
    \frac{\alpha_s \, C_F}{4\pi} \,
    \left[-2 + 2 \, L \right]
  \right) \right) +\frac{\alpha_s \, C_F}{4\pi} \, \Delta A_2,
\\[0.5em]
T_1 &=& \xi_\perp \, \left(1 +
    \frac{\alpha_s \, C_F}{4\pi} \,
    \left[  \ln \frac{m_b^2}{\mu^2} - L \right]
  \right)
+\frac{\alpha_s \, C_F}{4\pi} \, \Delta T_1,
\\[0.5em]
T_2 &=& \frac{2E}{M} \, \xi_\perp \, \left(1 +
    \frac{\alpha_s \, C_F}{4\pi} \,
    \left[ \ln \frac{m_b^2}{\mu^2} - L \right]
  \right)
+\frac{\alpha_s \, C_F}{4\pi} \, \Delta T_2,
\\[0.5em]
T_3 &=& \xi_\perp \,  \left(1 +
    \frac{\alpha_s \, C_F}{4\pi} \,
    \left[ \ln \frac{m_b^2}{\mu^2} - L \right]
  \right) 
\\&& 
   - \xi_\parallel \, \left(1 +
    \frac{\alpha_s \, C_F}{4\pi} \,
    \left[ \ln \frac{m_b^2}{\mu^2} -2  + 4 \, L \right]
  \right) +\frac{\alpha_s \, C_F}{4\pi} \, \Delta T_3
\label{vertexcorr2}
\eeq
for the remaining form factors of vector mesons. 
Here we introduced the abbreviation
\beq
  L &=& - \frac{2E}{M-2E} \, \ln\frac{2E}{M}
\label{Labbrev}
\eeq
with $L \to 1$ for $E \to M/2$ ($q^2 \to 0$). In defining $L$ 
we identified the $b$ quark mass $m_b$ and the $B$ meson mass 
$M$, since the difference is a $1/M$ effect.

The form factors receive a further additive correction from the 
interaction with the spectator quark, indicated by 
$\Delta F_i$ in Eqs.~(\ref{add1})-(\ref{vertexcorr2}). 
This correction will be 
calculated in the next subsection, see Eqs.~(\ref{eq:pi_final}), 
(\ref{hardrho0}), (\ref{hardrho}) below. The tensor form factors 
$f_T$, $T_{1,2,3}$ are scale-dependent,  
since the defining tensor currents are not conserved. 

\subsection{Hard spectator interaction}

\subsubsection{General considerations}

As discussed above a further correction at order $\alpha_s$ 
arises from the spectator 
interaction shown in Figs.~\ref{fig1}c and d. 
To leading order in $1/M$ we can restrict ourselves to
the two-particle light-cone distribution amplitudes of 
the $B$ meson and the light meson. 

The momenta of the $b$ quark and the spectator antiquark in the 
$\bar{B}$ meson are chosen as 
\begin{equation}
 p_b^\mu = m_b v^\mu,  \qquad  
    l^\mu  = \frac{l_+}{2} n_+^\mu + l_\perp^\mu 
               + \frac{l_-}{2} n_-^\mu, 
\label{momenta1}
\end{equation}
respectively, and 
\beq
&&  k_1^\mu = u E n_-^\mu + k_\perp^\mu + \frac{\vec k_\perp^2}{4uE} \,
n_+^\mu , \qquad
  k_2^\mu = \bar u E n_-^\mu - k_\perp^\mu + \frac{\vec k_\perp^2}{4\bar
uE} \, n_+^\mu
\label{momenta2}
\eeq
denote the momenta of the quark and antiquark in the light meson.  
As usual we have defined $\bar u = 1-u$. Note that all components 
of the spectator momentum $l$ are of order $\Lambda_{\rm QCD}$, while 
$k_{1,2}$ are of order $M$ along the $n_-$-direction with transverse 
components of order $\Lambda_{\rm QCD}$. 

The contribution of Figs.~\ref{fig1}c and d  
to the heavy-to-light current matrix elements
is now given by the convolution formula
\beq
  \langle L | \bar q \, \Gamma \, b |B\rangle^{(\rm HSA)} 
&=& \frac{4\pi\alpha_s C_F}{N_C} \, \int_0^1 du \int_0^\infty dl_+ \,
        M^B_{\beta\gamma} \, M^L_{\delta\alpha} \, 
 {\mathcal T}^\Gamma_{\alpha\beta\gamma\delta}.
\label{HSA}
\eeq
Here $\Gamma$ denotes an arbitrary Dirac matrix in the
heavy-to-light current, and
$ {\mathcal T}^\Gamma_{\alpha\beta\gamma\delta}$
is the hard-scattering amplitude, to be calculated from
the Feynman graphs in Figs.~\ref{fig1}c and d. Dirac indices
$\alpha,\beta,\gamma,\delta$ are written explicitly, while
the colour trace has already been performed. 

The relevant non-perturbative bound state dynamics of the 
initial and final mesons is encoded in the two-particle light-cone 
projectors $M^B$ and $M^L$ (not to be confused with the matrices 
${\mathcal M}_B$ and ${\mathcal M}_L$ 
used in Eq.~(\ref{IW-pi})). The expressions for these 
projectors are obtained after Fourier transformation to 
momentum space of the light-cone expansion of matrix elements of 
quark-antiquark operators. For light pseudoscalar and vector 
mesons the relevant details can be found in Ref.~\cite{Braun:1990iv} and 
Ref.~\cite{Ball:1998sk}, respectively, and the complete expressions are
summarised in Appendix~\ref{appa}. For pseudoscalar mesons with 
momentum $p'$ we have
\beq
  M_{\delta\alpha}^P &=& \frac{i \, f_P}{4} 
  \, \slash p' \gamma_5 \, \phi(u) + \ldots,
\label{pimeson}
\eeq
where $f_P$ is the pseudoscalar decay constant; for vector mesons
\beq
  M_{\delta\alpha}^V &=& - \frac{i}{4} \,
  \Bigg\{ f_\perp \,\slash\varepsilon^* \slash p' \, \phi_\perp(u) 
         +f_V \,  \slash p' \, \frac{m_V}{E}  (v\cdot\varepsilon^*)
  \,   \phi_\parallel(u) + \ldots \Bigg\}_{\delta\alpha}
\label{rhomeson}
\eeq
with $f_V$ and $f_\perp$ denoting the longitudinal and transverse 
vector meson decay constants, defined through
\begin{equation}
\langle V(p',\varepsilon^*)|\bar{q}\gamma_\mu q'|0\rangle 
= -i f_V m_V \varepsilon^*_\mu,
\qquad
\langle V(p',\varepsilon^*)|\bar{q}\sigma_{\mu\nu} q'|0\rangle 
= f_\perp (p'_\mu\varepsilon^*_\nu-p'_\nu\varepsilon^*_\mu).
\end{equation}
The ellipses in Eqs.~(\ref{pimeson}), (\ref{rhomeson}) denote 
twist-3 two-particle contributions as specified in  
Appendix~\ref{appa}.

The light-cone projectors for heavy mesons have not yet been discussed
in full generality in the literature. As shown in Appendix~\ref{appb} 
the projector we need is 
\beq
M_{\beta\gamma}^B &=&
-\frac{i f_B M}{4} \,
\left[\frac{1+\!\not\!v}{2} \left\{
\phi^B_+(l_+)\!\not\!n_++\phi^B_-(l_+)\left(\!\not \!n_- - l_+ 
\gamma_\perp^\nu\frac{\partial}{\partial l_\perp^\nu}\right)
\right\}\gamma_5\right]_{\beta\gamma} \, \Bigg|_{l=\frac{l_+}{2} n_+}.
\label{Bmeson}
\eeq
The derivative acts on the amplitude, expressed in terms of the
spectator quark momentum $l$, and subsequently $l$ is set equal 
to its plus-component. (In writing Eq.~(\ref{Bmeson}) we have 
assumed the relation (\ref{brel1}), which is valid only when one 
sets to zero the three-particle contributions to $\phi^B_-(l_+)$. 
However, keeping the more general form (\ref{bproj}) would not 
alter our result, since the symmetry-breaking correction turns 
out to involve only the distribution amplitude $\phi^B_+(l_+)$.)

In Feynman gauge, the hard-scattering amplitude is given by the 
expression
\beq
  {\mathcal T}^\Gamma_{\alpha\beta\gamma\delta}
  &=&
    \left[\Gamma \, \frac{m_b (1+\slash v) + \slash l - \slash k_2}
                       {(m_b v+l-k_2)^2-m_b^2} 
        \, \gamma_\mu
   + \gamma_\mu \, \frac{\slash k_1+\slash k_2- \slash l}
                           {(k_1+k_2-l)^2}
        \, \Gamma 
    \right]_{\alpha\beta} \, \frac{1}{(l-k_2)^2}\,
    \left[\gamma^\mu \right]_{\gamma\delta}
\nonumber\\
  &\simeq&
    \left[\Gamma \, \frac{m_b (1+\slash v) - \bar{u} E \slash n_-}
                       {4 \bar{u}^2 l_+ m_b E^2} 
        \, \gamma_\mu
   + \gamma_\mu \, \frac{E \slash n_- - \slash l}
                           {4\bar{u} l_+^2 E^2}
        \, \Gamma 
    \right]_{\alpha\beta} \, 
    \left[\gamma^\mu \right]_{\gamma\delta}.
\label{hsamplitude}
\eeq
To arrive at the second line we approximated the hard scattering 
amplitude by its leading term in the heavy quark limit, neglecting 
terms of order $\Lambda_{\rm QCD}/M$. The numerator of the 
second term, $E \slash n_- - \slash l$, is an exception to this. 
Here the subleading term $\slash l$ 
has to be kept, since the leading term is annihilated by the 
leading-twist light-cone projection operators: 
$\gamma^\mu M^{P,V}\gamma_\mu  \slash n_- = 0$. As a consequence 
both terms in the second line of Eq.~(\ref{hsamplitude}) 
are of order $1/(M\Lambda_{\rm QCD})$ in the heavy quark limit. 

The denominator of the first term vanishes quadratically 
with $\bar u$. Assuming, as usual, that
the leading-twist light-cone distribution amplitudes of the light
mesons vanish only linearly with $\bar u$, 
the term $m_b(1+\slash v)$ in the numerator generates a contribution
that diverges logarithmically for $\bar u \to 0$. Hence the amplitude 
is dominated by small gluon virtualities, i.e.\ soft physics. 
In order to justify our
factorisation ansatz (\ref{fff}), we have to show that the soft 
(endpoint dominated) contributions
do not break the heavy quark/large recoil symmetries and can be 
accounted for by a redefinition of 
$\xi_{P,\perp,\parallel}$. After this redefinition the 
hard scattering term $\Phi_B \otimes T_i \otimes \Phi_P$ in 
(\ref{fff}) is free of soft (endpoint) singularities and 
can be computed consistently in the hard scattering approach. 
To demonstrate this, we insert  
$1=(\slash n_+\slash n_-)/4 +(\slash n_- \slash n_+)/4$ 
in front of $\Gamma$ in the trace 
\begin{eqnarray}
&&\mbox{tr}(M^L \Gamma (1+\slash v) \gamma^\mu M^B\gamma_\mu) 
= 2\,\mbox{tr}(\slash v M^L\Gamma M_B)
\nonumber\\
&&\hspace*{0.5cm}
=\,\frac{1}{2} \mbox{tr}(\slash v M^L\slash n_+\slash n_- \Gamma M_B) 
+ \frac{1}{2} \mbox{tr}(\slash v M^L\slash n_-\slash n_+ \Gamma M_B).
\label{ttr}
\end{eqnarray}
The first term is already of the form of Eq.~(\ref{trace-form}), 
and therefore respects the heavy quark/large recoil symmetry
relations, while the second term with $\slash n_-$ next to 
$M_L$ vanishes. This shows that the endpoint singularities can 
be factorised into the soft form factors as necessary for the validity
of the factorisation formula (\ref{fff}). Consequently, the
symmetry-breaking corrections are due to large-momentum transfer
interactions. 

However, a further check needs to be performed. Some of the 
twist-3 distribution amplitudes that contribute to $M^L$ 
do not vanish as $\bar u \to 0$ and generate terms that diverge
{\em linearly}\/ at the endpoint. The linear divergence 
produces an additional factor $M/\Lambda_{\rm QCD}$ that compensates 
the $1/M$ suppression of the twist-3 distribution amplitudes.
Therefore, twist-3 distribution amplitudes contribute at leading
order to the soft part of the form factors.
In order to justify our factorisation ansatz, we have to show 
that these endpoint contributions can also be absorbed into 
a renormalisation of 
$\xi_{P,\perp,\parallel}$. Furthermore, the expression $E \slash n_-$ 
in the second term of (\ref{hsamplitude}) is {\em not} annihilated 
by the twist-3 part of the projector $M^L$, and this provides 
another source of {\em leading power} contributions. This 
contribution must also be shown to satisfy the symmetry relations. 
The first of these two contributions can be treated in a manner 
analogous to Eq.~(\ref{ttr}). The second term in the second line 
of (\ref{ttr}), which does not have the required symmetry-preserving 
structure, does not vanish, when twist-3 terms are included 
in $M^L$, but it turns out to be of order $\bar{u}$ as 
$\bar{u}\to 0$, hence not leading to the power-counting 
breaking linear divergence. Indeed, for a pseudoscalar meson, 
we obtain 
\begin{equation}
  M^P\slash n_-\slash n_+ = -\frac{i f_P}{4}\,\mu_P\gamma_5
 \left(\phi_p+\frac{\phi'_\sigma}{6}\right) 
 \slash n_-\slash n_+ + \ldots,
\label{comb1}
\end{equation}
where the ellipsis denote the contributions from distribution
amplitudes which vanish as $\bar{u}\to 0$. The important point is 
that while $\phi_p$ and $\phi'_\sigma$ do not vanish at the 
endpoint, the combination of both which appears in Eq.~(\ref{comb1}) 
does. Similarly, we obtain 
\begin{eqnarray}
  M^V\slash n_-\slash n_+ &=& -\frac{i m_V}{4} \bigg[
  f_V(\slash\varepsilon^*-(v\cdot\varepsilon^*) \slash n_-) 
  \left(g_\perp^{(v)}+\frac{1}{2} g_\perp^{(a)'}\right) 
\nonumber\\
&&\hspace*{-1cm}+ \,
  f_\perp \frac{m_V}{E}  (v\cdot\varepsilon^*) 
  \left(h_\parallel^{(t)}+\frac{1}{2} h_\parallel^{(s)'}\right) \bigg] \,  
 \slash n_-\slash n_+ + \ldots,
\label{comb2}
\end{eqnarray}
which also vanishes at $\bar{u}=0$ (see Appendix~\ref{appa}). 
This shows that all 
power-counting breaking linearly divergent contributions 
can be accounted for by a renormalisation of 
$\xi_{P,\perp,\parallel}$. The second of the two contributions 
discussed above is easily seen to preserve the heavy quark/large
recoil symmetries, since 
\begin{equation}
\mbox{tr}(M_L\gamma^\mu \slash n_-\Gamma M_B\gamma_\mu) 
= \mbox{tr}(\gamma_\mu M_L\gamma^\mu \slash n_-
\slash n_+\slash n_-\Gamma M_B) 
\end{equation}
has already the structure required by Eq.~(\ref{trace-form}). 

Before concluding this general discussion, we note that only the 
distribution amplitude $\phi^B_+(l_+)$ enters the {\em symmetry 
breaking} (but computable) corrections. This is seen immediately 
for the first term in Eq.~(\ref{hsamplitude}) since only 
the term $-\bar{u} E \slash n_-$ is relevant for the symmetry
breaking terms. The $\slash n_-$ annihilates the corresponding 
term in $M^B$, and since there is no dependence on $l_\perp$, 
$\Phi_-(l_+)$ does not contribute. In the case of the second term 
of Eq.~(\ref{hsamplitude}) we need to examine the $\slash l$ 
term. The term proportional to $\slash n_-$ in the $B$-meson
projector~(\ref{Bmeson}) drops out since the factor $\slash n_-$ can be 
anticommuted in such a way that it annihilates with
$M^L_{\delta\alpha}$. But the derivative term in Eq.~(\ref{Bmeson})
survives and  leaves a transverse Dirac matrix
$\gamma_\perp^\nu$ to the left of the matrix $\Gamma$. 
Inserting $1=(\slash n_+\slash n_-)/4 +(\slash n_- \slash n_+)/4$, 
the ``wrong'' projector $(\slash n_-\slash n_+)/4$ is annihilated by
the leading twist terms in $M^L$, since 
$\gamma_\perp^\nu$ anti-commutes with $\slash n_{\pm}$. The 
remaining terms involving $\Phi_-(l_+)$ then preserve the 
heavy quark/large recoil symmetries.

To summarise this subsection, the hard-scattering contributions,
calculated from Eq.~(\ref{HSA}) fall into two classes: 
soft contributions that formally diverge at the endpoints
$\bar u \to 0$ (or $l_+ \to 0$) but obey the symmetry relations
predicted by Eq.~(\ref{trace-form}), and hard contributions that
show regular behaviour at the endpoint and (potentially) break
the symmetries. These general considerations will be verified 
by the explicit calculation that follows below. 

\subsubsection{$B\to P$ form factors}

The hard-scattering contributions to the current matrix elements  
for $B\to P$ transitions are calculated using 
Eqs.~(\ref{HSA}), (\ref{hsamplitude}) together with the 
light-cone projection operators. The form factors are then 
determined by comparing the result with the
definitions in Eqs.~(\ref{fvector}) and (\ref{ftensor}). 

To give an example, the result for the form factor $f_+$ reads
\beq
  f_+^{(\rm HSA)} &=& \frac{\alpha_s C_F}{4\pi} \, \frac{ \pi^2 f_B f_P
    M}{ N_C E^2}
   \, \int_0^1 du \, \int_0^\infty dl_+ \, \Bigg\{
           \frac{4E-M}{M} \, \frac{\phi(u) \, \phi^B_+(l_+)}{\bar u l_+}
\\[0.2em]
&& \hspace*{-1cm}
 +\,\frac{(1+\bar u) \, \phi(u) \, \phi^B_-(l_+)}
                {\bar u^2 l_+} 
    +  
           \frac{ \mu_P}{2E} \,
    \left[ 
           \frac{ (\phi_p(u) - \phi^\prime_\sigma(u)/6) \, \phi^B_+(l_+)}
                {\bar u^2 l_+} + \frac{4 E \phi_p(u) \, \phi^B_+(l_+)}
                {\bar u l_+^2}\right]
\Bigg\}
\nonumber \label{fplusresult}
\eeq
Similar expressions are obtained for $f_0$ and $f_T$.
The three terms in the second line of Eq.~(\ref{fplusresult}) 
have endpoint singularities for $\bar u \to 0$ and/or
$l_+ \to 0$. As discussed in the previous subsection
these terms preserve the symmetry structure predicted by
Eq.~(\ref{trace-form}). In our factorisation scheme we can
absorb these contributions into $\xi_P$ without having to
specify a regularisation procedure for the endpoint
singularities. In fact, due to the renormalisation convention 
Eq.~(\ref{constraint}) the entire correction displayed in 
Eq.~(\ref{fplusresult}) is absorbed into $\xi_P$. The 
important point is that terms with the structure of those in the
second line (which cannot be computed with standard hard scattering methods) 
then do not appear in the other two form factors 
$f_0$ and $f_T$, whereas a term with the structure of that of the
first line (which is computable) does, as will be seen below.

It is worth emphasizing again that all four terms in 
Eq.~(\ref{fplusresult}) 
are of the same order $(\Lambda_{\rm QCD}/M)^{3/2}$ 
with respect to the $1/M$ power counting. 
In the current context this reflects the observation 
made in Ref.~\cite{Charles:1998dr} that in the QCD sum rule 
calculation some of the twist-3 distribution amplitudes 
contribute at leading power to the soft part of the form
factor. (We can make contact with the result of 
Ref.~\cite{Charles:1998dr} by expanding the light-cone distribution 
amplitudes around $\bar u =0$. Identifying 
\begin{equation}
I_2 \leftrightarrow
\frac{\alpha_s C_F}{4\pi} \, \frac{2  \pi^2 f_B^2M}{N_C} \,
\int\frac{du}{\bar u} \, \int \frac{dl_+}{l_+} \, \phi^B_-, 
\qquad I_1 \leftrightarrow
\frac{\alpha_s C_F}{4\pi} \, \frac{2\pi^2  f_B^2 M E}{N_C} \,
\int\frac{du}{\bar u^2} \, \int \frac{dl_+}{l_+} \, \phi^B_+,
\end{equation}
and using the endpoint behaviour of the distribution amplitudes, 
we reproduce the result quoted in Eq.~(87) of Ref.~\cite{Charles:1998dr}.)
We also note that the soft contributions involve the $\phi^B_-$ 
amplitude of the $B$-meson.

The first term in Eq.~(\ref{fplusresult}) represents a genuine hard 
scattering correction which is dominated by gluon virtualities of  
order $\mu^2 \simeq 2 E l_+\sim M\Lambda_{\rm QCD}$. At this order 
in $\alpha_s$, we need only two particular moments of the distribution
amplitudes to compute this correction. For the light pseudoscalar 
meson, we need 
\begin{equation}
\langle \bar{u}^{-1}\rangle_P = \int du \frac{\phi(u)}{\bar u}.
\end{equation}
This is the same moment of the leading twist distribution 
amplitude that contributes to the 
$P\gamma$ transition form factor (see e.g.\
Ref.~\cite{Feldmann:1999uf} and references therein). For the 
$B$ meson, we need
\begin{equation}
\langle l_+^{-1}\rangle_+=\int dl_+ \frac{\phi^B_+(l_+)}{l_+}. 
\label{bwfmom}
\end{equation} 
The same moment is also needed in $B \to \ell\nu\gamma$ 
decays~\cite{Korchemsky:1999qb}, and it also determines the
leading non-factorisable hard spectator corrections to 
$B\to\pi\pi$ decays \cite{Beneke:1999br,Beneke:1999br2}. 
It is convenient to define the quantity
\beq
\Delta F_P
    &=& \frac{8 \pi^2  f_B f_P}{N_C M} 
          \, \langle l_+^{-1} \rangle_+ \,
          \langle \bar u^{-1} \rangle_{P}. 
\label{pihard}
\eeq
The theoretical uncertainties in the computation of the hard scattering 
correction due to the moments defined above and the $B$ meson decay 
constant are all contained in this quantity. 

With the help of this notation, we now present the result for 
the hard scattering correction to 
$B \to P$ form factors, as defined by $\Delta f_{+,0,T}$ in
Eqs.~(\ref{add1}), (\ref{vertexcorr1}). The renormalisation 
convention~(\ref{constraint}) implies $\Delta f_+ \equiv  0$ 
by definition. The other two quantities are then given by 
\beq
&&
  \Delta f_0  =  \frac{M-2 E}{2 E} \, \Delta F_P,  \qquad
  \Delta f_T  = - \frac{M+m_P}{2 E}  \, \Delta F_P \ .
\label{eq:pi_final}
\eeq
Note that $\Delta f_0$ vanishes at $q^2=0$ ($E=M/2$) as required on 
general grounds.

\subsubsection{$B \to V$ form factors}

The analysis of the hard-scattering corrections to form factors
for $B \to V$ transitions proceeds in the same way as for
$B \to P$ decays. For instance, 
the contribution to the form factors $A_0$ and
$T_1$ reads
\beq
  A_0^{(\rm HSA)} &=& 
    \frac{\alpha_s C_F}{4\pi} \, \frac{ \pi^2 f_B M}{ N_C E^2}
    \, \int_0^1 du \int_0^\infty dl_+ \,
    \Bigg\{\frac{f_V \phi_\parallel(u) \, \phi^B_+(l_+)}{\bar u l_+}
    + \frac{(1+\bar u) \, f_V \phi_\parallel(u) \, \phi^B_-(l_+)}
      {\bar u^2 l_+}     
 \nonumber \\[0.2em]
 && \hspace*{-1cm} + \,\frac{m_V f_\perp}{2 E} \, 
\Bigg[ \frac{(-2 E)\, h_\parallel'{}^{(s)}(u)\,\phi^B_+(l_+)}
{\bar u l_+^2} + 
\frac{(h_\parallel^{(t)}(u) - h_\parallel'{}^{(s)}(u)/2)\, 
\phi^B_+(l_+)}{\bar u^2 l_+ }\Bigg]\Bigg\} 
\label{A0result} 
\\[0.2em]
 T_1^{(\rm HSA)} &=&   \frac{\alpha_s C_F}{4\pi} \, \frac{ \pi^2 f_B
    M}{ N_C E^2}
   \, \int_0^1 du \int_0^\infty dl_+ \,
  \Bigg\{
           \frac{2E}{M} \, \frac{f_\perp \phi_\perp(u) \, 
\phi^B_+(l_+)}{\bar u l_+}
 +
           \frac{f_\perp  \phi_\perp(u) \, \phi^B_-(l_+)}
                {\bar u^2 l_+}   
\\[0.2em]
 && \hspace*{-1cm} + \,\frac{m_V f_V}{2 E} \, 
\Bigg[ \frac{2 E\, (g_\perp^{(v)}(u) - 
g_\perp'{}^{(a)}(u)/4)\,\phi^B_+(l_+)}{\bar u l_+^2} + 
\frac{(g_\perp^{(v)}(u) - g_\perp'{}^{(a)}(u)/4) \,
  \phi^B_+(l_+)}{\bar u^2 l_+ }\Bigg]
\Bigg\} 
\nonumber \label{T1result}
\eeq
The last three terms in $A_0$ and $T_1$ represent again a universal
soft contribution that can be absorbed into $\xi_\parallel$ and
$\xi_\perp$, respectively. Only the first term gives rise to the
symmetry-breaking correction. We introduce the quantities 
\beq
\Delta F_\parallel
    &=& \frac{8 \pi^2  f_B f_V}{N_C M} 
          \, \langle l_+^{-1} \rangle_+ \,
          \langle \bar u^{-1} \rangle_{\parallel},
\nonumber\\[0.1em] 
\Delta F_\perp
    &=& \frac{8 \pi^2  f_B f_{\perp}}{N_C M} 
          \, \langle l_+^{-1} \rangle_+ \,
          \langle \bar u^{-1} \rangle_{\perp},
\label{rhohard}
\eeq
where $\langle \bar{u}^{-1}\rangle_\parallel = \int du \,
\phi_\parallel(u)/\bar u$, $\langle \bar{u}^{-1}\rangle_\perp = \int du\, 
\phi_\perp(u)/\bar u$. The renormalisation convention 
(\ref{constraint}) implies no correction to $A_0$ and $V$. 
The hard correction to the other 
$B\to V$ form factors, defined in
Eqs.~(\ref{add2})-(\ref{vertexcorr2}), 
reads
\beq
&& \hspace*{-0.5cm} \Delta A_1 = 0 , \qquad
  \Delta A_2 = \frac{m_V}{E}\,\frac{M}{M-m_V} \, \frac{M(M-2 E)}{4
    E^2} \, \Delta F_\parallel \ ,
\label{hardrho0}\\[0.2em]
&& \hspace*{-0.5cm} \Delta T_1= \frac{M}{4 E}\,\Delta F_\perp, \qquad
    \Delta T_2= \frac{1}{2} \, \Delta F_\perp, \qquad
    \Delta T_3= \frac{M}{4 E}\,\Delta F_\perp + \frac{m_V}{E}\, 
\left(\frac{M}{2 E}\right)^2\,\Delta F_\parallel.
\label{hardrho}
\eeq

\subsection{Summary of corrections
to form factor ratios}

The complete result for the form factors at order $\alpha_s$ 
is given by Eqs.~(\ref{add1})-(\ref{vertexcorr2}), together
with the expressions for the hard scattering correction in the 
previous subsection. 

We summarise here how the form factor 
ratios (\ref{pirelation})-(\ref{rhorelation2}) are modified 
by the sym\-metry-breaking corrections at order $\alpha_s$. 
For the $B\to$ pseudoscalar meson form factors we obtain
\beq
 \frac{f_0}{f_+} &=& \frac{2E}{M}   
  \left(1 +
    \frac{\alpha_s  C_F}{4\pi}  
    \left[ 2 - 2   L \right]  + \frac{\alpha_s C_F}{4\pi}
    \, \frac{M (M-2E)}{(2E)^2} \, \frac{\Delta F_P}{\xi_P} \right),
\label{corr1}
\\[0.2em]
 \frac{f_T}{f_+} &=&  
    \frac{m_P+M}{M}    \left(1 +
    \frac{\alpha_s  C_F}{4\pi}  
    \left[ \ln \frac{m_b^2}{\mu^2} + 2   L \right]
-\frac{\alpha_s C_F}{4\pi} \, \frac{M}{2E} \, \frac{\Delta F_P}{\xi_P} 
\right),
\eeq
The form factors for transitions between a $B$ meson and
transversely or longitudinally polarized light vector mesons
satisfy the relations 
\begin{equation}
  \frac{A_1}{V} = \frac{2 EM}{(M+m_V)^2},
\label{a1rel}
\end{equation}
\begin{equation}
  \frac{T_1}{V} = \frac{M}{2E} \,
  \frac{T_2}{V} = \frac{M}{M+m_V}   \left(
    1 + \frac{\alpha_s C_F}{4\pi}   \left[ \ln\frac{m_b^2}{\mu^2} -
      L\right] 
      + \frac{\alpha_s C_F}{4\pi} \, \frac{M}{4E} \, 
\frac{\Delta F_\perp}{\xi_\perp}
   \right),
\label{corr2}
\end{equation}
and
\begin{eqnarray}
 \frac{(M+m_V)/(2E) \, A_1 - (M-m_V)/M \,
    A_2}{(m_V/E)\,A_0} &=&
\nonumber\\
&&\hspace*{-6cm}
   1 + \frac{\alpha_s C_F}{4\pi}  
    \left[-2+2L\right]
    -\frac{\alpha_s C_F}{4\pi} \, \frac{M(M-2E)}{(2 E)^2} \, \frac{\Delta
      F_\parallel}{(E/m_V)\,\xi_\parallel},
\end{eqnarray}
\begin{equation}
  \frac{(M/2E) \, T_2 - T_3}{(m_V/E) A_0} =
  1 + \frac{\alpha_s C_F}{4\pi}  
     \left[ \ln \frac{m_b^2}{\mu^2} - 2 + 4 L \right]
   - \frac{\alpha_s C_F}{4\pi} \, \left(\frac{M}{2E}\right)^2 \frac{\Delta
      F_\parallel}{(E/m_V)\,\xi_\parallel},
\label{corr3}
\end{equation}
respectively. The various quantities $\xi_{P,\perp,\parallel}$,
$\Delta F_{P,\perp,\parallel}$ and $L$ are defined in 
Eqs.~(\ref{constraint}), (\ref{pihard}),(\ref{rhohard}) and 
(\ref{Labbrev}). The radiatively corrected form factor ratios 
(\ref{corr1})-(\ref{corr3}) constitute the main result of
this paper. Note that two relations, namely the ones between
$A_1$ and $V$ and between $T_1$ and $T_2$ do not receive 
$\alpha_s$ corrections to leading order in the $1/M$
expansion.

\section{Numerical analysis}

We now turn to the numerical analysis of the form factor 
ratios~(\ref{corr1})-(\ref{corr3}). We take a pion as a 
representative pseudoscalar meson and a $\rho$ meson as a 
representative vector meson. There is little theoretical 
uncertainty in the evaluation of the vertex correction. Assuming 
that the scale-dependent tensor form factors are renormalized 
at the scale $\mu_1=m_b$, the only uncertainty arises from the 
scale of $\alpha_s$, which we also take to be $\mu_1$. 

The hard scattering correction is more difficult to estimate. Although
it depends only on universal quantities, some of them, $\langle
l_+^{-1}\rangle_+$ in particular, are not well known. Note 
that the characteristic scale for the hard scattering correction 
is $(M\Lambda_{\rm QCD})^{1/2}$ and all quantities in the 
hard scattering correction are evaluated at $\mu_2=1.47\,$GeV. 
We use the following input parameters:

{\em Meson decay constants.} The pion decay constant 
$f_\pi=131\,$MeV. For the $\rho$ meson decay constant 
we assume  $f_\rho=198\,$MeV and 
$f_{\rho,\perp}(\mu_2)=152\,$MeV, coincident with 
Ref.~\cite{Ball:1998kk}. (We ignore a small difference due to the 
slightly different renormalisation scale.) Finally, 
$f_B=180\,$MeV. The uncertainty in the evaluation of the 
hard scattering correction coming from the decay constant 
is estimated to be around $\pm 15\%$.

{\em Light-cone distribution amplitudes.} Since 
$\langle l_+^{-1}\rangle_+\sim 1/\Lambda_{\rm QCD}$, but nothing else is
known about this parameter at present, we estimate it to 
be $(0.2-0.5\,\mbox{GeV})^{-1}$ and take $(0.3\,\mbox{GeV})^{-1}$ 
as our central value. (Using the $B$ meson distribution amplitude 
suggested in Ref.~\cite{Grozin:1997pq}, one obtains 
$\langle l_+^{-1}\rangle_+ = (0.32\,\rm GeV)^{-1}$.) The situation is 
more favourable for the light mesons. The asymptotic
distribution amplitude $\phi_\pi(u) = 6 u \bar u$ is now 
favored by the CLEO data \cite{Gronberg:1998fj} on the 
$\pi\gamma$ form factor (see also
Refs.~\cite{Kroll:1996jx,Feldmann:1999uf}), so that 
$\langle \bar u^{-1}\rangle_\pi =3$. For the $\rho$ meson we 
use the result for the second Gegenbauer moment quoted in 
Ref.~\cite{Ball:1998kk}, which leads to 
$\langle \bar u^{-1}\rangle_\parallel =3.48$ and 
$\langle \bar u^{-1}\rangle_\perp = 3.51$. The uncertainty in the 
evaluation of the 
hard scattering correction coming from the meson distribution 
amplitudes is estimated to be around $\pm 50\%$ with most of the uncertainty 
due to the $B$ meson. 

{\em Soft form factors.} We also need an estimate for the
absolute value of the functions $\xi_{\pi}$, $\xi_\perp$
and $\xi_\parallel$ at large recoil in order to compute ratios 
such as $\Delta F_P/\xi_P$. Given the conditions (\ref{constraint}), 
we use $f_+(q^2=0)$, $V(0)$ and $A_0(0)$ as input and parametrise the 
energy dependence by the energy dependence of the soft form 
factor in the heavy quark/large recoil limit, Eq.~(\ref{softbpi}), 
i.e.\ we take 
\begin{equation}
\left\{\xi_P,\frac{M+m_V}{M}\,\xi_\perp,\frac{E}{m_V}\,\xi_\parallel\right\} = 
\left\{f_+(0),V(0),A_0(0)\right\} \times \left(\frac{M}{2 E}\right)^2 
\end{equation}
with
\begin{equation}
\left\{f_+^\pi(0),V^\rho(0),A_0^\rho(0)\right\} = 
\left\{0.305,0.338,0.372\right\}
\end{equation}
from the QCD sum rule calculations of 
Refs.~\cite{Ball:1998kk,Ball:1998tj}. The uncertainty in the 
evaluation of the 
hard scattering correction coming from the soft form factor 
is estimated to be around $\pm 20\%$. 

\begin{figure}[p]
\begin{center}
   \vspace{0cm}
   \epsfysize=5cm
   \epsfxsize=7.5cm
   \epsffile{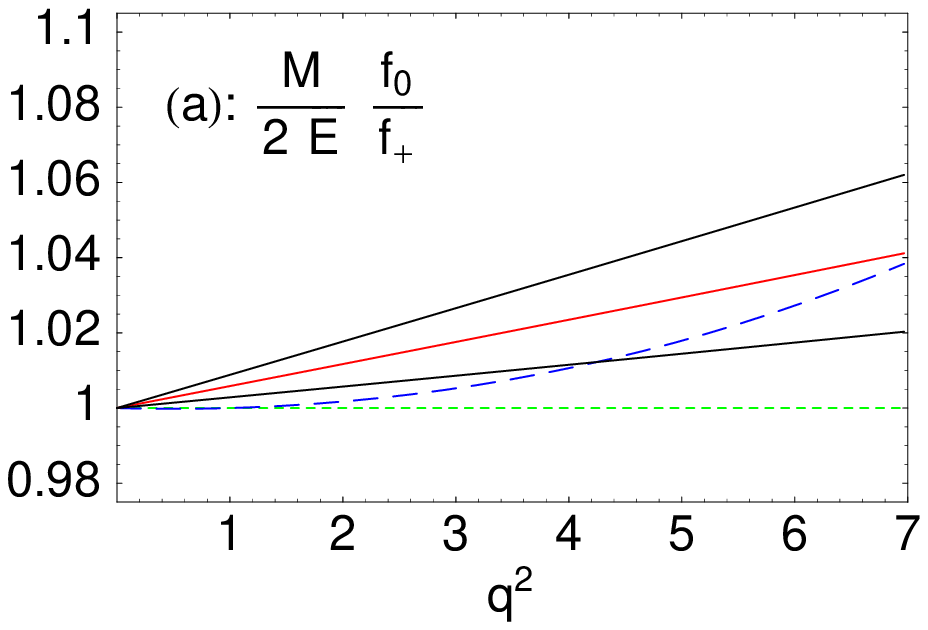}
   \epsfysize=5cm
   \epsfxsize=7.5cm
   \epsffile{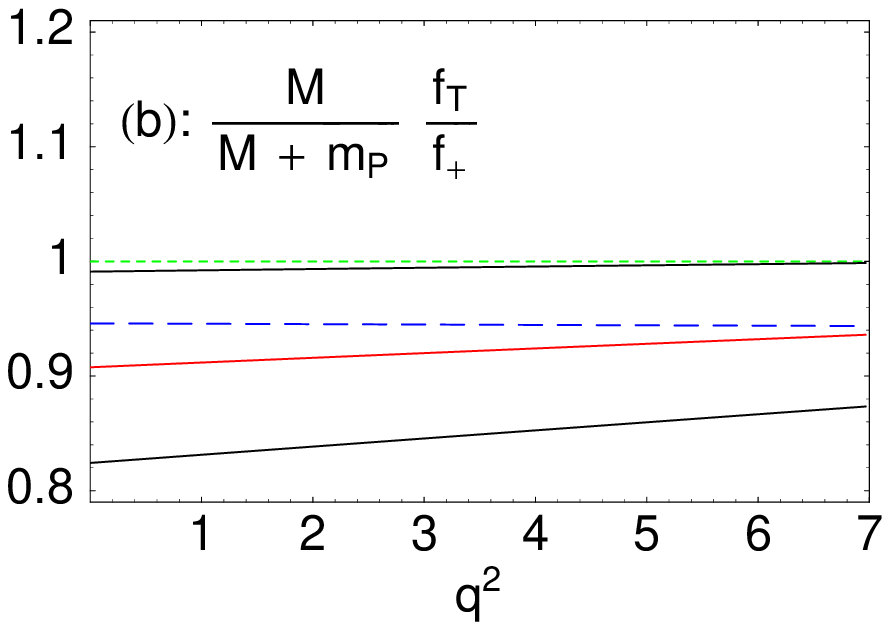}
   \epsfysize=5cm
   \epsfxsize=7.5cm
   \epsffile{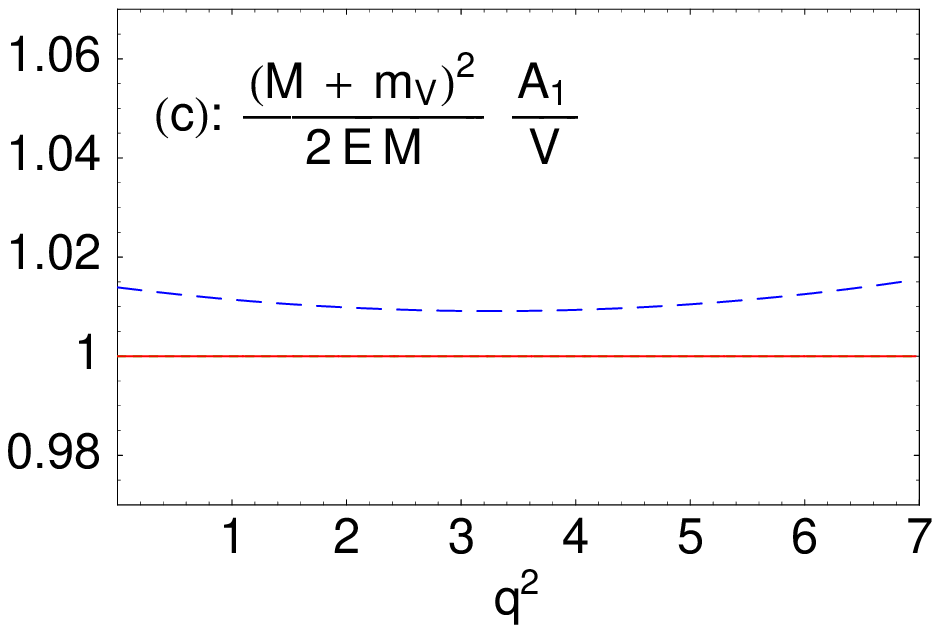}
   \epsfysize=5cm
   \epsfxsize=7.5cm
   \epsffile{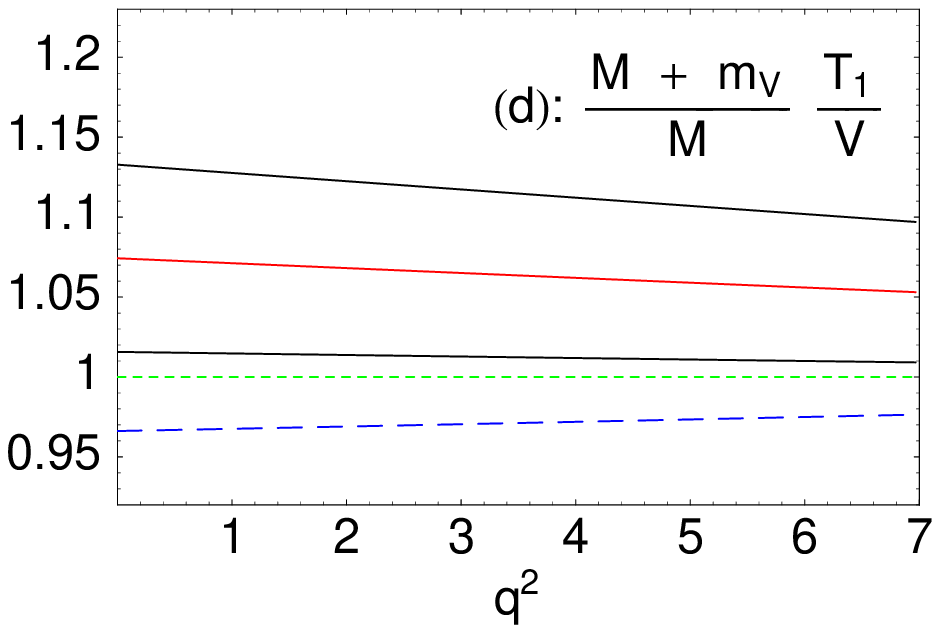}
   \epsfysize=5cm
   \epsfxsize=7.5cm
   \epsffile{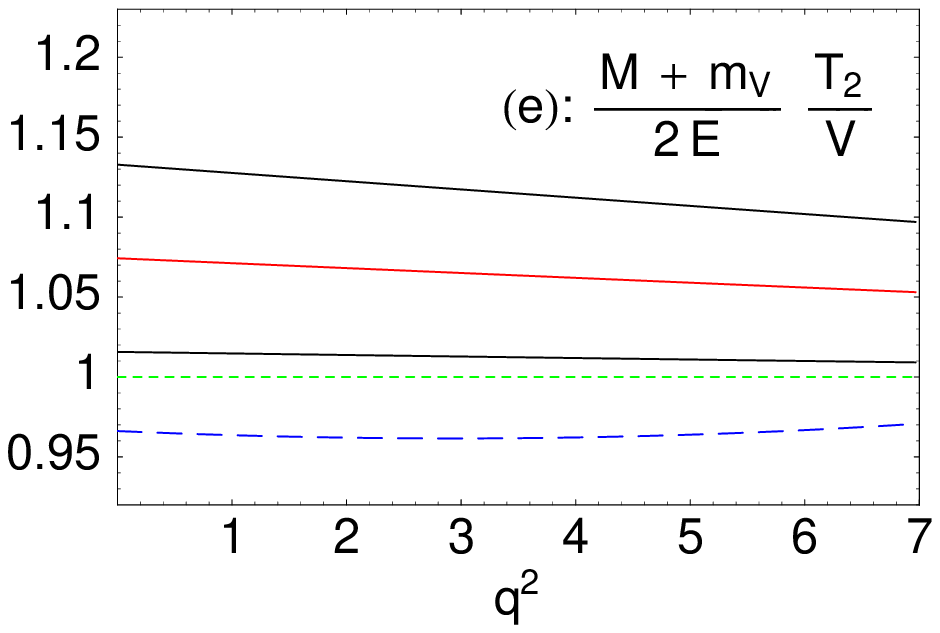}
   \epsfysize=5cm
   \epsfxsize=7.5cm
   \epsffile{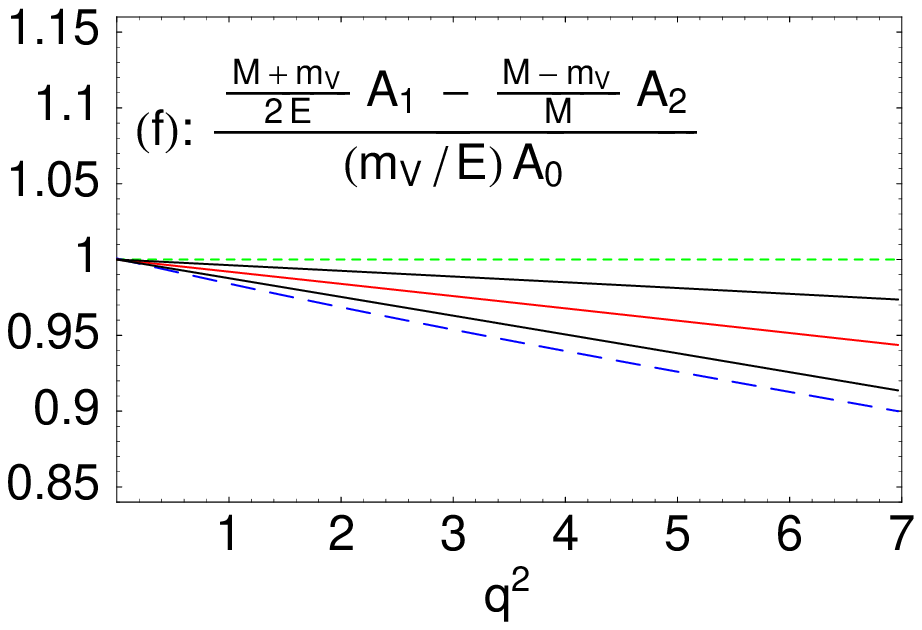}
   \epsfysize=5cm
   \epsfxsize=7.5cm
   \epsffile{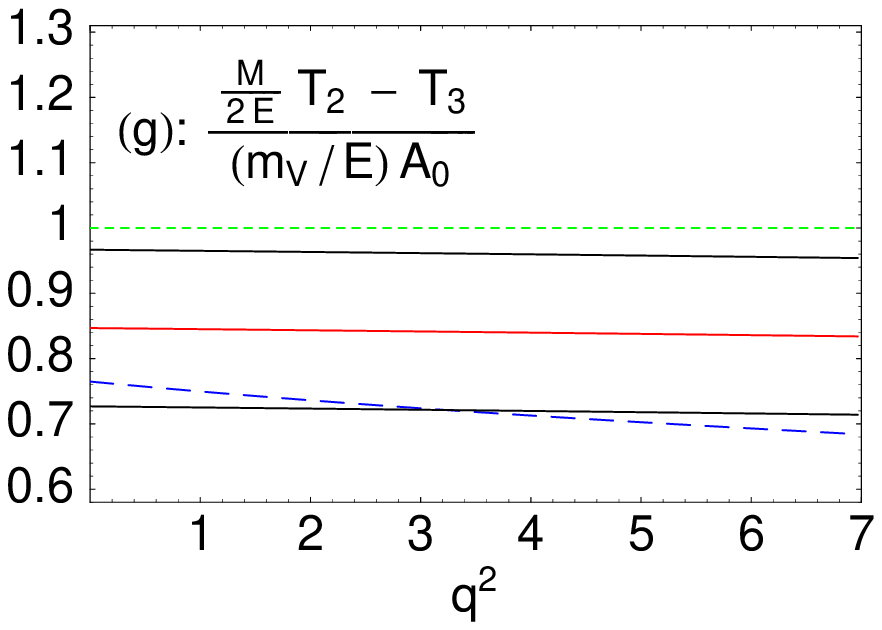}
   \vspace{-0.6cm}
\end{center}
\caption[dummy]{\label{fig2}\small 
Corrections to symmetry relations as a
function of $q^2$ (three solid lines, see text).
Figs.~(a)--(g) correspond to the form factor ratios
in Eqs.~(\ref{corr1})-(\ref{corr3}) with overall scaling
factors divided out, such that the symmetry limit (dotted lines) corresponds 
to a ratio equal to 1 independent on $q^2$. 
For illustration we show also results from
 QCD sum rules \cite{Ball:1998tj,Ball:1998kk} (dashed lines).
Tensor form factors are renormalized at $\mu=m_b$.}
\end{figure}

Combining these input parameters, we obtain 
\begin{equation}
\label{vall}
\frac{\Delta F_P^\pi}{f_+^\pi(0)} =3.85,\qquad
\frac{\Delta F_\perp^\rho}{V^\rho(0)} =4.72,\qquad
\frac{\Delta F_\parallel^\rho}{A_0^\rho(0)} =5.54
\end{equation}
with an overall uncertainty of about $60\%$. The result for 
the form factor ratios (\ref{corr1})-(\ref{corr3}) is shown as 
the central solid curve in Fig.~\ref{fig2}, using 
$\alpha_s(\mu_1)=0.22$ and $\alpha_s(\mu_2)=0.34$. The other two solid 
curves follow from multiplying the numbers in Eq.~(\ref{vall}) 
by 0.4 and 1.6, respectively, and reflect the current 
theoretical uncertainty in evaluating the symmetry 
breaking correction. As a general rule, the hard scattering correction
is larger than the vertex correction, and therefore the 
theoretical uncertainty remains significant. A determination 
of the $B$ meson parameter $\langle l_+^{-1}\rangle_+$ would be 
very helpful to eliminate the single most important theoretical 
uncertainty. The typical size of the symmetry-breaking correction 
is of the order of $10\%$ in the range $0\leq q^2 \leq M^2/4$. 
An exception is the last relation (g),
between $T_2$, $T_3$ and $A_0$, which might receive a negative correction 
of $30\%$. Since the large recoil symmetries apply only for 
small $q^2 \ll M^2$, the $q^2$ range shown is restricted to values
smaller than $7\,\mbox{GeV}^2$.

It is instructive to compare our results with the 
QCD sum rule calculations. In Fig.~\ref{fig2} 
we plot the effective form factor parametrisation for 
$B\to \pi$ transitions from Ref.~\cite{Ball:1998tj} and 
for $B \to \rho$ transitions from Ref.~\cite{Ball:1998kk}. 
In general we find a fair agreement of the two results, as 
far as the magnitude and sign of the symmetry-breaking 
correction is concerned, with the exception of the ratios 
$T_1/V$ and $T_2/V$ (Fig.~\ref{fig2}d and e). 
In particular, we note that in both
approaches the relation between $A_1$ and $V$
(Fig.~\ref{fig2}c) and the relation between $T_1$
and $T_2$ (Fig.~\ref{fig2}d vs.\ Fig.~\ref{fig2}e)
receive practically no corrections. The QCD sum rule calculations 
also include some $1/M$ corrections, as well as quadratic meson 
mass effects, while the ratios (\ref{corr1})-(\ref{corr3}) 
are strictly valid at leading order in the heavy quark 
expansion. The two calculations are therefore not directly 
comparable at the level of $1/M$ effects and this may 
explain the remaining numerical differences. 

The advantage of the present approach over the QCD sum rule approach
is that it does not require the duality assumption in the $B$ meson 
channel, but refers directly to the light-cone distributions of the 
$B$ meson. This is allows us to compute radiative corrections to the 
symmetry limit with less model-dependence than in the sum rule 
approach. The price for this is that the result depends on the 
parameter $\langle l_+^{-1}\rangle_+$, which may perhaps be 
constrained in the future, but remains poorly known for now. 
Furthermore, the duality assumption in the QCD sum rule method 
permits the calculation of the soft contribution to the form 
factors, which must be taken as an input, when one relies on symmetries
and hard scattering methods only.

\section{Application to the forward-backward asymmetry in
$B \to V \ell^+ \ell^-$}

The forward-backward (FB) asymmetry in the decay 
$B \to V \ell^+ \ell^-$ (where $V$ is a vector meson, for example 
a $K^*$ meson) provides an interesting example, where the model-independent 
form factor ratios derived above may be useful. Burdman 
\cite{Burdman:1998mk} noted, using form factor models, that the 
location of the forward-backward asymmetry zero was nearly 
independent on the form factor models he considered. An explanation 
of this fact was given by Ali {\em et al.} \cite{Ali:1999mm}, 
who noted that the form factor ratios on which the asymmetry 
zero depends are predicted free of hadronic uncertainties in 
the heavy quark/large energy limit considered in 
Ref.~\cite{Charles:1998dr}. We are now in the position to discuss 
the effect of radiative corrections to the symmetry limit. 

The decay $B \to K^* \ell^+ \ell^-$ is induced by the
flavour-changing neutral current transition $b \to s \ell^+ \ell^-$. 
In the Standard Model, after integrating
out top quarks, $W$ and $Z$ bosons, it is described by the 
effective weak hamiltonian reviewed in detail 
in Ref.~\cite{Buchalla:1996vs}. Let us begin the discussion by 
considering only the `semi-hadronic' operators in 
the effective hamiltonian
\begin{equation}
  {\mathcal H}_{\rm eff} =
  - \frac{G_F}{\sqrt2} \, V_{ts}^* \, V_{tb} \,
  \sum_{i=7,9,10} C_i(\mu) \, \mathcal {O}_i(\mu)
\label{Heff}
\end{equation}
where 
\begin{equation}
{\cal O}_7 =\frac{e}{8\pi^2}\,m_b\bar{s}\sigma^{\mu\nu}
(1+\gamma_5)b F_{\mu\nu},
\vspace*{0.2cm}
\end{equation}
\begin{equation}
{\cal O}_9 = [\bar{s}\gamma^\mu(1-\gamma_5) b] 
[\bar{l}\gamma_\mu l],
\qquad
{\cal O}_{10} = [\bar{s}\gamma^\mu(1-\gamma_5) b] [\bar{l}\gamma_\mu 
\gamma_5 l],
\nonumber 
\end{equation}
and $C_i(\mu)$ are the corresponding short-distance Wilson
coefficients. The Wilson coefficients are calculable in the 
standard model, but they may also receive
contributions from new particles in theories beyond the SM, 
and hence their experimental determination is of great 
interest. (Of course extensions of the standard model may introduce a 
larger set of operators as well.) The FB asymmetry zero in 
the decay $B \to K^* \ell^+ \ell^-$ provides a way to determine 
$C_9$. 

The forward-backward asymmetry is defined as 
\begin{equation}
\frac{dA_{\rm FB}}{dq^2} = \int_0^1 d(\cos\theta) \,\frac{d^2\Gamma}{dq^2 
d\cos\theta} - \int_{-1}^0 d(\cos\theta) \,\frac{d^2\Gamma}{dq^2 
d\cos\theta},
\end{equation}
where $\theta$ is the angle between the positively charged lepton and
the $B$ meson in the $\ell^+\ell^-$ pair rest frame, and $q^2$ is the 
invariant mass of the lepton pair. The double-differential decay 
width is obtained by computing the matrix elements 
$\langle K^* \ell^+\ell^-|{\cal O}_{7,9,10}|\bar{B}\rangle$. The hadronic 
part of the matrix elements is parametrised in terms of the 
form factors in Eqs.~(\ref{V})-(\ref{ffdef}). Without going into 
the details here (which can be found in Ref.~\cite{Ali:1991is,Ali:1999mm} or 
elsewhere), the FB asymmetry is found to be 
\begin{eqnarray}
\frac{dA_{\rm FB}}{dq^2} &\propto& C_{10} \Bigg[C_9 V(q^2) A_1(q^2) + 
\frac{m_b}{q^2} C_7\Big(V(q^2) T_2(q^2) (M-m_{K^*})
\nonumber\\ 
&&\hspace*{2cm}+\,
A_1(q^2) T_1(q^2) (M+m_{K^*})\Big)\Bigg],
\end{eqnarray}
If this vanishes for a certain value $s_0=q_0^2$, then 
\beq
  C_9 &=& - \frac{m_b}{s_0} \, C_7
  \left\{ \frac{T_2(s_0)}{A_1(s_0)} \, (M-m_{K^*})
         + \frac{T_1(s_0)}{V(s_0)} \, (M+m_{K^*}) \right\}
\label{relation}
\eeq
Making use of Eqs.~(\ref{a1rel}), (\ref{corr2}), we find that 
\begin{equation}
\frac{T_2(s_0)}{A_1(s_0)} \, (M-m_{K^*}) =  
\frac{T_1(s_0)}{V(s_0)} \, (M+m_{K^*}),
\end{equation}
including radiative corrections (neglecting, as always, terms of order 
$m_{K^*}^2/M^2$), so that Eq.~(\ref{relation}) can 
be written as
\beq
  C_9
  &=&  - \frac{2 M m_b}{s_0} \, C_7
   \, \left(1 + 
  \frac{\alpha_s C_F}{4\pi} \, \left[ \ln \frac{m_b^2}{\mu^2} - L \right] 
        +  \frac{\alpha_s C_F}{4\pi} \, 
 \frac{\Delta F_\perp}{\xi_\perp(s_0)}
\right).
\label{relation2} 
\eeq

Consider first the leading order result. Even at leading order 
one cannot neglect the effect of four quark operators and the 
chromomagnetic dipole operator in the weak effective hamiltonian. 
Their effect is conventionally taken into account by defining 
``effective'' Wilson coefficients, such that $C_7\to C_7^{\rm eff}$ 
and $C_9 \to \mbox{Re}(C_9^{\rm eff}(s_0))$ in 
Eq.~(\ref{relation2}). $C_9^{\rm eff}(s_0)$ is not a true 
short-distance quantity, acquiring $q^2$-dependence and an 
imaginary part, which turns Eq.~(\ref{relation2}) into an 
implicit equation for $s_0$. Assuming standard model values for 
the Wilson coefficients, the solution is 
$s_0=2.9\,\mbox{GeV}^2$ \cite{Ali:1999mm}, 
which is small enough to justify the 
application of large recoil symmetries. The magnitude of 
the radiative correction to the symmetry limit of 
Eq.~(\ref{relation2}) can be deduced from Fig.~\ref{fig2}d to 
be $(6.5\pm 5)\%$. This range can be considerably 
narrowed, when more information on the moment (\ref{bwfmom}) of the 
$B$ meson distribution amplitude becomes available. Since 
$C_7^{\rm eff}$ is already constrained to be close to its 
standard model value from the
measurement of inclusive $b \to s \gamma$ decays, 
Eq.~(\ref{relation2}) 
provides an almost model-independent determination 
of $C_9$ as soon as $s_0$ is measured, and assuming that 
the matrix elements of the four quark operators can be 
computed to sufficient accuracy. (For completeness we note 
that the form factor ratios required for the present analysis 
are exactly those where there exists a discrepancy with the 
QCD sum rule result. From Fig.~\ref{fig2}d, we also deduce that 
the QCD sum rule calculation of the relevant
form factor ratios leads to a 3\% reduction of $s_0$. 
This adds importance to clarifying the origin of this 
discrepancy.)

\begin{figure}[t]
   \vspace{-3.2cm}
   \epsfysize=25.2cm
   \epsfxsize=18cm
   \centerline{\epsffile{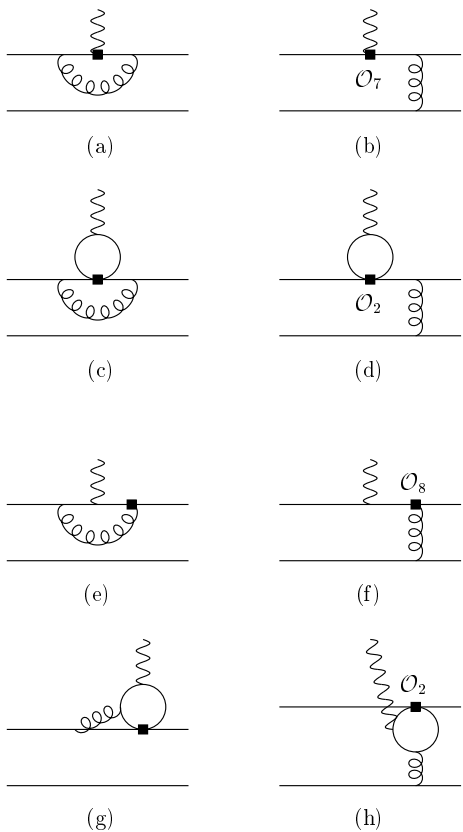}}
   \vspace*{-10.8cm}
\caption[dummy]{\label{fig3}\small Various next-to-leading 
order contributions to the $B\to K^*\gamma^*$ matrix elements.}
\end{figure}

We should emphasize that Eq.~(\ref{relation2}) is not a complete 
result at order $\alpha_s$ even after replacing the Wilson 
coefficients $C_{7,9}$ by effective coefficients. A technical way to 
see this is to note that the $\mu$-dependent logarithm in 
Eq.~(\ref{relation2}), which arises from the scale-dependent 
tensor form factor, does not compensate completely the 
renormalisation scale dependence of $C_7^{\rm eff}$. There exist 
further corrections at order $\alpha_s$, originating from 
four quark operators and the chromomagnetic dipole operator in the 
weak effective hamiltonian, which cannot be 
expressed in terms of form factors, i.e.\ matrix elements of the type 
$\langle K^*|\bar{s}\Gamma b|\bar{B}\rangle$. 
Sample Feynman diagrams are shown in Fig.~\ref{fig3}e-g, 
compared to the diagrams in Fig.~\ref{fig3}a-d, which do assume 
the structure of form factor matrix elements. 

However, drawing upon the factorisation formula for 
{\em non-leptonic} $B$ decays \cite{Beneke:1999br,Beneke:1999br2},  
we note that the matrix elements of {\em all} operators in 
the weak effective hamiltonian, including four quark operators, 
can be expressed as 
\begin{equation}
\langle K^* \ell^+\ell^-|{\cal O}_i|\bar{B}\rangle = C_i \, \xi  + 
\Phi_B \otimes T_i \otimes \Phi_{K^*},
\end{equation}
i.e.\ in a form similar to Eq.~(\ref{fff}). This allows us to 
compute the corrections of the type shown in Fig.~\ref{fig3}e-g 
without introducing further non-perturbative 
parameters and to discuss exclusive radiative and semi-leptonic 
decays in a systematic way, comparable to the case of 
form factors and 
non-leptonic $B$ decays. This extension of the present work, and 
a complete discussion of the radiatively corrected FB asymmetry 
zero, will be presented elsewhere.

\section{Conclusion}

In this article we reconsidered the heavy quark/large recoil 
symmetries for heavy-to-light $B$ meson form factors (at large 
recoil) discussed first in Ref.~\cite{Charles:1998dr}. We find 
that these symmetries, discussed originally for the soft parts 
of the form factors, survive radiative corrections in the sense 
that symmetry-breaking effects are dominated by hard scattering 
and therefore computable with standard methods. The structure 
of the corrections is tentatively summarised by the factorisation 
formula Eq.~(\ref{fff}), which is similar to the factorisation 
formula for non-leptonic $B$ decays 
\cite{Beneke:1999br,Beneke:1999br2}. However, 
much more theoretical work is needed to establish 
Eq.~(\ref{fff}), or a variant thereof, beyond the calculation 
of symmetry-breaking effects.

The symmetries, and the calculation of symmetry-breaking effects, 
put the discussion of heavy-light form factors at large recoil 
on a similar conceptual footing as heavy-light form factors at small
recoil or heavy-heavy form factors. In the symmetry limit 
(heavy quark limit) the three independent form factors for $B\to P$ 
transitions reduce to a single function $\xi_P$ for each 
pseudoscalar meson, and the seven independent form factors  
for $B \to V$ transitions reduce to two functions 
$\xi_\perp$, $\xi_\parallel$ for each vector meson, 
corresponding to transverse or longitudinal polarization of the  
vector meson. Symmetry-breaking effects come from hard 
gluon corrections and fall into two classes: 
vertex corrections to the heavy-to-light current, which can
be treated in an analogous way as in heavy quark effective
theory, and hard rescattering with the spectator quark which is described 
by the hard-scattering approach and which involves light-cone
distribution amplitudes of the participating mesons. The second class 
of corrections is a specifically new element of form factors 
at large recoil.

The numerical evaluation of the symmetry-breaking corrections
typically yields 10\% effects to form factor ratios; larger effects 
are possible for some form factor ratios, while two ratios do not 
receive any correction at order $\alpha_s$. There is at present 
a sizeable uncertainty in evaluating the hard-scattering correction, 
which seems to limit the usefulness of the present analysis. 
However, the major part of this uncertainty is due to a single 
moment of the $B$ meson distribution amplitude. This moment is 
a $B$-meson-universal quantity, and since only recently it has 
been realised that this quantity appears in many 
$B$ decays, we should expect this quantity to be determined 
much more accurately in the future. 

The form factor relations at large recoil turn out to be strikingly 
useful for the forward-backward asymmetry in 
the exclusive decay $B\to K^* \ell^+\ell^-$, as already noted 
in Ref.~\cite{Ali:1999mm}. Here we find symmetry-breaking corrections 
on the order of 5\%, but a definite conclusion must await the 
calculation of non-form factor type corrections at order 
$\alpha_s$. We plan to complete this task in a future publication. 

\section*{Acknowledgements}

We would like to thank V.M.~Braun, G.~Buchalla and G.P.~Korchemsky for
helpful discussions.

\begin{appendix}

\section{Light-cone projections of light mesons}
\label{appa}

\subsection{Twist-2 and -3 distribution amplitudes of pseudoscalars}

We follow the conventions of Ref.~\cite{Braun:1990iv} in defining 
the light-cone distribution amplitudes of light pseudoscalar mesons: 
\beq
  \langle P(p')|\bar q(y) \gamma_\mu \gamma_5 q(x)|0\rangle
  &=& - i  f_P \,p'_\mu \, \int_0^1 du \,  e^{i (u \, p'\cdot y +
    \bar u p' \cdot x)}\, \phi(u),
\nonumber \\[0.1em]
  \langle P(p')|\bar q(y) i \gamma_5 q(x)|0\rangle
  &=& f_P \mu_P \,  \int_0^1 du \, 
  e^{i (u \, p'\cdot y +\bar u \, p' \cdot x)} \, \phi_p(u),
\nonumber \\[0.1em]
  \langle P(p')|\bar q(y) \sigma_{\mu\nu}\gamma_5 q(x) |0\rangle
  &=& i f_P \mu_P \, (p'_\mu z_\nu - p'_\nu z_\mu) \,
      \int_0^1 du \, e^{i (u \, p'\cdot y +\bar u \, p' \cdot x)} \,
      \frac{\phi_\sigma(u)}{6}.
\label{pidadef}
\eeq
Here we defined $z=y-x$, and $f_P$ is the decay constant. The 
parameter $\mu_P=M_P^2/(m_1+m_2)$ is proportional to the chiral 
condensate. (This definition holds for charged mesons, in which case 
the flavours of the light quarks with masses $m_1$ and $m_2$ are 
different.) $\phi(u)$ is the leading twist-2 distribution amplitude. All 
three distribution amplitudes are normalised to 1, as can be
seen by taking the limit $x \to y$. Above and 
in the following we implicitly assume that matrix elements are supplied with
the appropriate path-ordered exponentials of gluon fields in
order to make the definitions of distribution amplitudes 
gauge-invariant. The above definitions can be combined into 
\beq
&&  \langle P(p')|\bar q_\alpha(y) \, q_\delta(x)|0\rangle
= \frac{i f_P}{4} \, \int_0^1 du \,  e^{i (u \, p'\cdot y +
    \bar u p' \cdot x)} 
\nonumber\\[0.1cm]
&& \hspace*{2.5cm}\times\,\left\{ \slash p' \gamma_5 \, \phi(u)
                 - \mu_P \gamma_5 \left( \phi_p(u)
                 - \sigma_{\mu\nu} p'{}^{\mu} 
                 z^\nu \, \frac{\phi_\sigma(u)}{6}\right)
\right\}_{\delta\alpha}
\label{pispacedef}
\eeq

To obtain the momentum space projector $M^L_{\delta\alpha}$, which 
appears in Eq.~(\ref{HSA}), we take the Fourier transform 
of Eq.~(\ref{pispacedef}), using
\beq
  z^\lambda &\to&  (-i)\frac{\partial}{\partial k_{1\lambda}}  
   = (- i) \left\{\frac{n_+^\lambda}{2E} \,
  \frac{\partial}{\partial u} +\, \frac{n_-^\lambda}{2} \,
  \frac{\partial}{\partial k_1^-} + \frac{\partial}{\partial
    k_\perp{}_\lambda}\right\} 
\label{derivative}
\eeq
under Fourier transformation, 
where $k_1$ is the quark momentum defined in  Eq.~(\ref{momenta2}). 
The light-cone vectors $n_\pm$ are defined as after Eq.~(\ref{leet}) 
and the transverse projection is defined with respect to these two 
vectors. The derivative is supposed to act on the hard 
scattering amplitude in the momentum space representation. 
The derivative with respect to the momentum fraction $u$ can be 
made to act on the light-cone distribution amplitude by 
partial integration; the second term in Eq.~(\ref{derivative})
drops out; the third term, which involves the derivative with 
respect to the transverse momentum, acts on the hard-scattering
amplitude before the collinear limit $k_1=u p'=u E n_-$ is taken.
The light-cone projection operator of light pseudoscalar mesons 
in momentum space, including twist-3 two-particle contributions, 
then reads
\begin{equation}
M_{\delta\alpha}^P = \frac{i \, f_P}{4} 
  \, \Bigg\{ \slash p' \gamma_5 \, \phi(u)  - 
   \mu_P \gamma_5 \left(  \phi_p(u) 
       - i \sigma_{\mu\nu}\, n_-^\mu v^\nu \,
       \frac{\phi^\prime_\sigma(u)}{6}
 +i \sigma_{\mu\nu}  p'{}^\mu \, \frac{\phi_\sigma(u)}{6}
       \, \frac{\partial}{\partial
         k_\perp{}_{\nu}} \right)
 \Bigg\}_{\delta\alpha}.
\label{pimeson2}
\end{equation}
A complete description of the pseudoscalar meson at the twist-3 level 
would also include three-particle quark-antiquark-gluon 
contributions, which we do not give here 
(see Ref.~\cite{Braun:1990iv}). 

The asymptotic limit of the leading twist distribution amplitude 
is $\phi(u)=6 u \bar{u}$. The twist-3 two-particle distribution 
amplitudes are completely determined by the three-particle
distributions via the equations of motions except for a single 
term. In the approximation that we set all three-particle
distributions to zero (which is not an approximation that can be 
justified in any limit, but which is nonetheless useful to gain 
some insight in the structure of twist-3 two-particle 
contributions), the two twist-3 distribution amplitudes $\phi_p$ and
$\phi_\sigma$ are related by the equations of motion
\begin{equation}
  \frac{u}{2} \left(\phi_p(u) + \frac{\phi^\prime_\sigma(u)}{6} \right) =
  \frac{\phi_\sigma(u)}{6}, 
\qquad
  \frac{\bar{u}}{2} \left(\phi_p(u) - \frac{\phi^\prime_\sigma(u)}{6} 
  \right) = \frac{\phi_\sigma(u)}{6}, 
\label{eompi}
\end{equation}
which imply $\phi_p(u)=1$ and $\phi_\sigma(u)=6 u \bar{u}$. Note 
that $\phi_p$ and $\phi^\prime_\sigma$ do not vanish at the 
endpoints.

\subsection{Twist-2 and -3 distribution amplitudes of vector mesons}

We follow Ref.~\cite{Ball:1998sk} in defining the light-cone 
distribution amplitudes of light vector mesons with the 
exception that our convention for the outgoing light vector meson 
state differs by a factor $(-i)$ from the one of 
Ref.~\cite{Ball:1998sk}. (This is necessary for consistency with 
the conventions used for the form factors in
Eqs.~(\ref{V})-(\ref{ffdef}).)
We restrict ourselves 
again to the twist-2 and twist-3 two-particle amplitudes (neglecting 
terms suppressed by $m_V^2/E^2$ whenever they are not multiplied by
the longitudinal polarisation vector), and define $z=y-x$ with 
$z^2=0$ as before. To correctly account for meson mass effects, 
we define the meson momentum to be $P'$ with ${P'}^2=m_V^2$ and 
the light-like vector $p'_\mu=P'_\mu-m_V^2 z_\mu/(2 P'\cdot z)$.  
The chiral-even amplitudes are given by 
\begin{equation}
  \langle V(P',\varepsilon^*)|\bar q(y) \gamma_\mu q(x)|0\rangle
  = - i  f_V m_V \, \int_0^1 
      du \,  e^{i (u \, p'\cdot y +
    \bar u p' \cdot x)}
   \left\{p^\prime_\mu \,
    \frac{\varepsilon^*\cdot z}{p'\cdot z} \, \phi_\parallel(u)
         +\varepsilon_\perp^*{}_\mu \, g_\perp^{(v)}(u) \right\},
\label{rhoevendef0}
\end{equation}
\begin{equation}
  \langle V(P',\varepsilon^*)|\bar q(y) \gamma_\mu\gamma_5 q(x)|0\rangle
  = i  f_V m_V \,\epsilon_{\mu\nu\rho\sigma} \, 
      \varepsilon^*{}^\nu p^{\prime\rho} z^\sigma \,
    \int_0^1 du \,  e^{i (u \, p'\cdot y +
    \bar u p' \cdot x)} \,  
       \frac{g_\perp^{(a)}(u)}{4},
\label{rhoevendef}
\end{equation}
with $f_V$ being the usual vector meson decay constant.
The chiral-odd light-cone distribution amplitudes are given by
\beq
   \langle V(P',\varepsilon^*)|\bar q(y) \sigma_{\mu\nu} q(x)
            |0\rangle
  &=& - f_\perp \,\int_0^1 du \, e^{i (u \, p'\cdot y +
    \bar u p' \cdot x)} \, 
\Bigg\{(\varepsilon^*_\perp{}_\mu p_{\nu}^\prime -
  \varepsilon_\perp^*{}_\nu  p^\prime_{\mu}) \, \phi_\perp(u) 
\nonumber\\
&& \hspace*{-2cm}
  + \,\frac{m_V^2\,\varepsilon^*\cdot z}{(p'\cdot z)^2} \, 
   (p^\prime_\mu z_\nu -
    p^\prime_\nu  z_\mu) \, h_\parallel^{(t)}(u)
\Bigg\}, 
\end{eqnarray}
\begin{equation}
  \langle V(P',\varepsilon^*)|\bar q(y) q(x)|0\rangle =
    - f_\perp \,m_V^2 \, \varepsilon^*\cdot z\, 
    \int_0^1 du \, e^{i (u \, p'\cdot y +
    \bar u p' \cdot x)}  \, 
          \frac{h_\parallel^{(s)}(u)}{2}.
\label{rhooddef}
\end{equation}
The chiral-odd amplitudes involve the (scale-dependent) 
transverse decay constant $f_\perp$. We have neglected terms 
proportional to the light-quark masses.
The  longitudinal and transverse projections of the polarization 
tensor are defined as \cite{Ball:1998sk}
\beq
&&  \varepsilon^*_\parallel{}_\mu \equiv
     \frac{\varepsilon^* \cdot z}{P'\cdot z} \left(
      P'_\mu-\frac{m_V^2}{P'\cdot z} \,z_\mu\right), \qquad
 \varepsilon^*_\perp{}_\mu
        = \varepsilon^*_\mu -\varepsilon^*_\parallel{}_\mu .
\eeq
Note that the longitudinal projection of the polarization vector
counts as ${\mathcal O}(E/m_V)$. 
Eqs.~(\ref{rhoevendef0})-(\ref{rhooddef}) can be combined into
the expression
\beq
  &&  \langle V(P',\lambda)|\bar q_\alpha(y) \, q_\delta(x)|0\rangle
= -\frac{i}{4} \, \int_0^1 du \,  e^{i (u \, p'\cdot y +
    \bar u p' \cdot x)} 
\nonumber\\[0.1cm]
  && \quad \times\,\left\{ f_V m_V \left(
    p^\prime_\mu \, \frac{\varepsilon^*\cdot z}{p'\cdot z} \,
    \phi_\parallel(u) +  \slash \varepsilon^*_\perp \,
    g_\perp^{(v)}(u) +  \epsilon_{\mu\nu\rho\sigma} \,
    \varepsilon^*{}^\mu  p^{\prime\rho} z^\sigma \, \gamma^\mu\gamma_5
    \, \frac{g_\perp^{(a)}(u)}{4}\right) \right.
\nonumber \\[0.1em]
  && \qquad \,\,\,\left. 
  + \,f_\perp \Bigg(\slash\varepsilon_\perp^* \slash p^\prime \,
  \phi_\perp(u) - i \,  
   \frac{m_V^2\,\varepsilon\cdot z}{(p'\cdot z)^2}
  \, \sigma_{\mu\nu} \,p^{\prime\mu} z^\nu
  \, h_\parallel^{(t)}(u) - i \, m_V^2 \, \varepsilon^*\cdot z \,
  \frac{h_\parallel^{(s)}(u)}{2} \Bigg) \right\}_{\delta\alpha}\!\!.
\eeq
To perform the Fourier transform we first express the previous equation 
in terms of the $z$-independent vectors $P'$ and $\epsilon^*$. 
The Fourier transform of the terms with $p'\cdot z=P'\cdot z$ in the 
denominator can be treated by partial integration, leading 
to integrals over the distribution amplitudes.
Potential surface terms vanish as a consequence of the relations
(\ref{rel2}) and (\ref{rel4}) below (provided that additional gluon
contributions vanish as well). After the Fourier transform is taken 
we introduce the two light-like vectors $n_\pm$, and write 
$P^\prime_\mu=E n_{-\mu}+m_V^2 n_{+\mu}/(4 E)$. The transverse plane 
is now defined with respect to the two vectors $n_\pm$.  
We then obtain for the 
momentum space representation of the vector meson light-cone
projection:
\begin{equation}
  M_{\delta\alpha}^V =  M_{\delta\alpha}^V{}_\parallel + 
                          M_{\delta\alpha}^V{}_\perp 
\label{rhomeson2}
\end{equation}
with
\beq
M^V_\parallel &=& -\frac{if_V}{4} \, \frac{m_V(\varepsilon^*\cdot
  n_+)}{2 E} \,E\,
 \slash n_- \,\phi_\parallel(u)
 -\frac{if_\perp m_V}{4}  \,\frac{m_V(\varepsilon^*\cdot n_+)}{2 E}
 \, \Bigg\{-\frac{i}{2}\,\sigma_{\mu\nu} \,  n_-^\mu  n_+^\nu \,
 h_\parallel^{(t)}(u) 
\nonumber\\[0.1cm]
&& \hspace*{-0.0cm}
- \,i E\, \int_0^u dv \,(\phi_\perp(v) - h_\parallel^{(t)}(v)) \ 
     \sigma_{\mu\nu}   n_-^\mu 
     \, \frac{\partial}{\partial k_\perp{}_\nu}
  +\frac{h_\parallel'{}^{(s)}(u)}{2}\Bigg\}\, \Bigg|_{k=u p'} 
\eeq
and
\beq
   M^V_\perp &=& -\frac{if_\perp}{4} \, 
   E\,\slash \varepsilon^*_\perp \slash n_- \, \phi_\perp(u)
 -\frac{if_Vm_V}{4} \,\Bigg\{\slash \varepsilon^*_\perp\, g_\perp^{(v)}(u)
\nonumber\\[0.1cm] && 
-   \,E \, \int_0^u dv\, (\phi_\parallel(v) - g_\perp^{(v)}(v))  \
       \slash n_- \, \varepsilon^*_{\perp\mu} \,\frac{\partial}{\partial
         k_{\perp\mu}}
\cr &&
+ \,i \epsilon_{\mu\nu\rho\sigma} \,
        \varepsilon_\perp^{*\nu} n_-^\rho\, \gamma^\mu\gamma_5 
         \left[ n_+^\sigma \,\frac{g_\perp'{}^{(a)}(u)}{8}-
          E\,\frac{g_\perp^{(a)}(u)}{4} \, \frac{\partial}{\partial
         k_\perp{}_\sigma}\right]
 \Bigg\}
 \, \Bigg|_{k=up'}, 
\eeq
and where now
\begin{equation}
\varepsilon_\perp^\mu \equiv \varepsilon^\mu - 
\frac{\varepsilon\cdot n_+}{2}\,n_-^\mu- 
\frac{\varepsilon\cdot n_-}{2}\,n_+^\mu .
\end{equation}
In the main body of the text we usually neglect power-suppressed 
higher-twist effects, i.e.\ we identify the meson momentum $P'$ with 
$p'\equiv E n_-$ and set $\varepsilon^*\cdot n_-=0$.

The twist-3 distribution amplitudes are related to
the twist-2 ones by Wandzura-Wilczek--type relations,
namely \cite{Ball:1998sk}
\beq
  g_\perp^{(v)}(u) &=& \frac12 \left[ \,\int_0^u 
    \frac{\phi_\parallel(v)}{\bar v} \, dv  + \int_u^1 
    \frac{\phi_\parallel(v)}{v}\, dv  \right] +\ldots ,
\nonumber \\[0.2em]
  g_\perp^{(a)}(u) &=& 2 \left[ \bar u \int_0^u 
    \frac{\phi_\parallel(v)}{\bar v}\, dv  + u \, \int_u^1 
    \frac{\phi_\parallel(v)}{v}\, dv  \right] +\ldots
\label{eomeven}
\eeq
for the chiral-even amplitudes, and
\beq
    h_\parallel^{(t)}(u) &=& (2u-1) \left[ \,\int_0^u 
    \frac{\phi_\perp(v)}{\bar v} \, dv  - \int_u^1 
    \frac{\phi_\perp(v)}{v}\, dv  \right]  +\ldots,
\nonumber \\[0.2em]
  h_\parallel^{(s)}(u) &=& 2 \left[ \bar u \int_0^u 
    \frac{\phi_\perp(v)}{\bar v}\, dv  + u \, \int_u^1 
    \frac{\phi_\perp(v)}{v}\, dv  \right] +\ldots
\label{eomodd2}
\eeq
for the chiral-odd amplitudes.
The ellipses in Eqs.~(\ref{eomeven}), (\ref{eomodd2}) (and the following ones) 
indicate additional contributions from three-particle distribution
amplitudes containing gluons and terms proportional to light quark 
masses, which we do not consider here.
Eqs.~(\ref{eomeven}), (\ref{eomodd2}) also imply
\beq
 \frac{g_\perp'{}^{(a)}(u)}{4} + g_\perp^{(v)}(u) = \int_u^1 
 \frac{\phi_\parallel(v)}{v} \, dv +\ldots
\label{rel1}
\eeq
\vspace*{-0.3cm} 
\beq 
  \int_0^u \, (\phi_\parallel(v)-g_\perp^{(v)}(v)) \, dv =
   \frac12 \left[ \bar u \int_0^u 
    \frac{\phi_\parallel(v)}{\bar v}\, dv  - u \, \int_u^1 
    \frac{\phi_\parallel(v)}{v}\, dv  \right] +\ldots,
\label{rel2}
\eeq 
and
\beq
 \frac{h_\parallel'{}^{(s)}(u)}{2} + h_\parallel^{(t)}(u) = 
  - 2 \bar u \left[ \,\int_0^u \frac{\phi_\perp(v)}{\bar v} \, dv -
  \int_u^1 
 \frac{\phi_\perp(v)}{v} \, dv \right] +\ldots, 
\label{rel3}
\eeq
\vspace*{-0.3cm} 
\beq 
  \int_0^u \, (\phi_\perp(v)-h_\parallel^{(t)}(v)) \, dv =
   u \, \bar u \, \left[ \,\int_0^u 
    \frac{\phi_\perp(v)}{\bar v}\, dv  - \int_u^1 
    \frac{\phi_\perp(v)}{v}\, dv  \right] +\ldots
\label{rel4}
\eeq 
Again all distribution amplitudes are normalized to unity.
$\phi_\perp$, $\phi_\parallel$, $g_\perp^{(a)}$, and $h_\parallel^{(s)}$
vanish at the endpoints, whereas $g_\perp^{(v)}$, and
$h_\parallel^{(t)}$ do not.

\section{Derivation of the $B$ meson projection}
\label{appb}

\subsection{The momentum space projector}

In this appendix we derive the result for the $B$ meson projection 
operator stated in Eq.~(\ref{Bmeson}) starting from the two-particle
light-cone matrix element in coordinate space. We follow the convention of 
Ref.~\cite{Grozin:1997pq} and introduce the two functions 
$\tilde \phi^B_\pm(t)$ through the Lorentz decomposition 
of the following light-cone matrix element: 
\begin{equation}   
\langle 0 | \bar q_\beta (z) \, P(z,0) \, b_\alpha(0) |
  \bar B(p) \rangle = - \frac{i f_B M}{4}
  \left[ \frac{1+\slash v}{2} \,
         \left\{ 2 \tilde \phi^B_+(t) + 
                 \frac{\tilde \phi^B_-(t)-\tilde \phi^B_+(t)}{t} 
                 \, \slash z \right\} \, \gamma_5 \right]_{\alpha\beta}.
\label{ansatz1}
\end{equation}
We assume that $z^2=0$, defined $t=v\cdot z$, $p=M v$ and the 
path-ordered exponential
\begin{equation}
P(z_2,z_1) = \mbox{P}\exp\left(i g_s\int\limits_{z_2}^{z_1} 
dz^\mu A_\mu(z)\right).
\end{equation}
Eq.~(\ref{ansatz1}) is the most general parametrisation compatible 
with Lorentz-invariance and the heavy quark limit. 
The prefactor is chosen in such a way that for $z=0$ 
one obtains
\beq
\langle 0 | \bar q_\beta \left[\gamma^\mu
  \gamma_5\right]_{\beta\alpha} b_\alpha |
  \bar B(p) \rangle 
&=&
 i f_B M \, v^\mu
\eeq
if $\tilde \phi^B_+(t=0) = \tilde \phi^B_-(t=0)=1$. 

Let us call $M(z)$ the matrix element in Eq.~(\ref{ansatz1}) and 
$A(z)$ ($A(l)$) the hard scattering amplitude in coordinate 
(momentum) space. Then we obtain the momentum space projector 
$M^B$ of Eq.~(\ref{Bmeson}) through the identity
\begin{equation}
\label{id}
\int d^4 z \,M(z) A(z)
= \int\frac{d^4 l}{(2\pi)^4}\,A(l)\int d^4 z\,e^{-i l z} M(z) 
\equiv \int_0^\infty dl_+\,M^B A(l)\,\Big|_{l=\frac{l_+}{2} n_+},
\end{equation}
with $l$ decomposed as in Eq.~(\ref{momenta1}), 
\begin{displaymath}
l^\mu=\frac{l_+}{2} n_+^\mu + \frac{l_-}{2} n_-^\mu+l_\perp^\mu.
\end{displaymath}
The factors $\slash z$ and $1/(v\cdot z)$ that appear in 
Eq.~(\ref{ansatz1}) can be removed by having a derivative act on the 
hard scattering amplitude, and by partial integration, as 
in the case of the light meson distribution amplitudes. If we then 
introduce the momentum space distribution amplitudes through 
\beq
\tilde \phi^B_\pm(t) &\equiv &
  \int_0^\infty d\omega \, e^{-i \omega t} \, \phi^B_\pm(\omega),  
\eeq
we obtain 
\begin{eqnarray}
\label{id2}
\int d^4 z \,M(z) A(z) &=& 
- \frac{i f_B M}{4}\Bigg[ \frac{1+\slash v}{2} \int_0^\infty d\omega 
\,\Bigg\{ 2\phi^B_+(\omega) 
\nonumber\\
&& \hspace*{-1.5cm}
-\,\int_0^\omega d\eta \,\left(\phi^B_-(\eta)-
\phi^B_+(\eta)\right)\,\gamma^\mu\frac{\partial}{\partial l_\mu}
\Bigg\} \, \gamma_5 \Bigg]_{\alpha\beta}\,A(l)_{\beta\alpha}\,
\Big|_{l=\omega v}.
\end{eqnarray}
This is close to the desired expression except that $l=\omega v = 
\omega (n_++n_-)/2$. However, the hard scattering amplitude $A(l)$ for 
a decay into an energetic light meson moving in the $n_-$ direction 
has the property that it is independent of $l_-$ at leading 
order in the heavy quark expansion. More precisely, it can be 
written as 
\begin{equation}
A(l) = A^{(0)}(l_+) + l_\perp^\mu A^{(1)}_\mu(l_+) + O(1/M).
\end{equation}
Hence the $n_-$ component of $v$ does not contribute and we may 
set $l=\omega n_+/2$, which amounts to identifying $\omega$ and 
$l_+$ in view of Eq.~(\ref{momenta1}). Using 
\begin{equation}
\frac{\partial}{\partial l_\mu} = n_-^\mu \frac{\partial}{\partial l_+}+
n_+^\mu \frac{\partial}{\partial l_-}+\frac{\partial}{\partial
  l_{\perp\mu}}, 
\end{equation}
(and dropping the derivative with respect to $l_-$), we obtain 
\begin{eqnarray}
\label{bproj}
M^B_{\beta\alpha} &=& 
- \frac{i f_B M}{4}\Bigg[ \frac{1+\slash v}{2} \Bigg\{
\phi^B_+(\omega)\,\slash n_+ +  \phi^B_-(\omega)\,\slash n_- 
\nonumber\\
&& \hspace*{2cm}
- \,\int_0^{l_+} d\eta \,\left(\phi^B_-(\eta)-
\phi^B_+(\eta)\right)\,\gamma^\mu\frac{\partial}{\partial l_{\perp\mu}}
\Bigg\} \, \gamma_5 \Bigg]_{\alpha\beta}.
\end{eqnarray}
It is understood that $l=l_+ n_+/2$ is set after performing the 
derivative. The $B$ meson light-cone projector assumes the form 
quoted in Eq.~(\ref{Bmeson}) after implementing the equation of 
motion constraint, Eq.~(\ref{brel1}), 
derived in the following subsection. 

Notice that 
the independence of the hard scattering amplitude on $l_-$ 
is exactly the property that guarantees that we need the bilocal 
matrix element (\ref{ansatz1}) {\em on the light-cone} ($z^2=0$).
 
\subsection{Equation of motion constraint} 

We shall now show that the equation of motion for the light 
spectator quark relates $\phi^B_-(l_+)$ to $\phi^B_+(l_+)$ and 
three-particle quark-antiquark-gluon distribution amplitudes. This 
is similar to what happens for the twist-3 two-particle amplitudes 
of light vector mesons. In the approximation that the three-particle 
amplitudes are set to zero, we can determine $\phi^B_-(l_+)$ in 
terms of $\phi^B_+(l_+)$. (These type of relations are sometimes 
referred to as ``Wandzura-Wilczek relations'' \cite{Wandzura:1977qf}.)

In order to derive this relation we employ the 
equation of motion for the light quark in Eq.~(\ref{ansatz1}).
Since the derivative with respect to $z_\mu$ has to be taken
before the limit $z^2 \to 0$ let us, for the moment,
extend the definitions in  Eq.~(\ref{ansatz1}) to the case
$z^2\neq 0$ via $\tilde \phi^B_\pm(t) \to \tilde \phi^B_\pm(t,z^2)$.
Requiring the right-hand side of  Eq.~(\ref{ansatz1}) to vanish after
application of $\left[\slash \partial_{z_2}\right]_{\beta\gamma}$
(which is true only if the three-particle Fock-state $b\bar q g$ 
is neglected), and requiring $\tilde \phi^B_\pm(t,z^2)$ to 
not vanish as $z^2\to 0$, we obtain 
\beq
  \frac{\partial \tilde \phi^B_-}{\partial t} +
    \frac{1}{t} \, (\tilde \phi^B_- - \tilde \phi^B_+)
  \, \Big|_{z^2=0} &=& 0,\label{ww1}
 \\
  \frac{\partial \tilde \phi^B_+}{\partial z^2} +
    \frac{1}{4} \, \frac{\partial^2 \tilde \phi^B_-}{\partial t^2}
    \,  \Big|_{z^2=0} &=& 0.
\label{wwaux}
\eeq 
The second equation  is uninteresting for our purpose.
The first equation gives the desired  
relation between $\tilde \phi^B_+$ and $\tilde \phi^B_-$
in coordinate space. In terms of the momentum space distribution 
amplitudes, Eq.~(\ref{ww1}) reads
\begin{equation}
\label{brel1}
\int_0^{l_+} d\eta \,\left(\phi^B_-(\eta)-
\phi^B_+(\eta)\right) = l_+ \phi^B_-(l_+)
\qquad\mbox{or}\qquad
 \phi^B_+(l_+) = - l_+  \, \phi^{\prime\,B}_-(l_+),
\end{equation}
which is solved by 
\begin{equation}
\phi^B_-(l_+) = \int\limits_0^1 
                    \frac{d\eta}{\eta}\,\phi^B_+(l_+/\eta).
\label{ww2}
\end{equation}
In terms of Mellin moments one has ($N\geq 1$)
\beq
  \langle l_+^{N-1} \rangle_+ &=& N,   \langle l_+^{N-1} \rangle_-
\ , \qquad \left[\langle l_+^{N-1} \rangle_\pm
\equiv\int\limits_0^\infty
dl_+ \, l_+^{N-1} \, \phi^B_\pm(l_+) \right].
\label{mm}
\eeq  
The relation~(\ref{mm}) for $N=2$ has been derived independently 
in Ref.~\cite{Grozin:1997pq} from the equations of motions 
for the {\em heavy quark}\/ and Lorentz invariance. This yields
$\langle l_+ \rangle_+ = 2 \langle l_+ \rangle_- = 4/3 \bar
\Lambda$, where $\bar \Lambda$ is the leading contribution to the
mass difference $M-m_b$ in HQET. For $N=3$ a similar analysis gives
\cite{Grozin:1997pq}
\beq
&&  \langle l_+^2\rangle _+ = 2 \bar \Lambda^2 + \frac{2
    \lambda_E^2+\lambda_H^2}{3}, \qquad 
  \langle l_+^2\rangle _- = \frac23 \bar \Lambda^2 + \frac{
    \lambda_H^2}{3}, 
\eeq
where $\lambda_E$ and $\lambda_H$ 
parametrise the contributions 
of the chromoelectric and chromomagnetic fields to
the mass difference $M - m_b$.
Note that the relation~(\ref{mm}) is again satisfied if
we set $\lambda_E=\lambda_H$, and in particular if both 
quantities vanish which is
equivalent to neglecting the three-particle Fock state as we have done. 
Grozin and Neubert \cite{Grozin:1997pq} have also proposed the 
simple model distribution amplitudes $\phi^B_+(l_+) = l_+/l_{+0}^2 \, 
\exp[- l_+/l_{+0}]$, $\phi^B_-(l_+) = 1/l_{+0} \, 
\exp[- l_+/l_{+0}]$ inspired by a QCD sum rule analysis.
It is easy to see that they satisfy the
relations~(\ref{ww2}), (\ref{mm}) exactly.

\end{appendix}


\begin{thebibliography}{99}

\bibitem{Beneke:1999br}
M. Beneke et~al.,
\newblock Phys. Rev. Lett. {\bf 83} (1999) 1914, hep-ph/9905312.

\bibitem{Beneke:1999br2}
M. Beneke et~al.,
\newblock hep-ph/0006124.

\bibitem{Charles:1998dr}
J. Charles et~al.,
\newblock Phys. Rev. {\bf D60} (1999) 014001, hep-ph/9812358.

\bibitem{Isgur:1989vq}
N. Isgur and M.B. Wise,
\newblock Phys. Lett. {\bf B232} (1989) 113.

\bibitem{Isgur:1990ed}
N. Isgur and M.B. Wise,
\newblock Phys. Lett. {\bf B237} (1990) 527.

\bibitem{Buchalla:1996vs}
G. Buchalla, A.J. Buras and M.E. Lautenbacher,
\newblock Rev. Mod. Phys. {\bf 68} (1996) 1125, hep-ph/9512380.

\bibitem{Ali:1999mm}
A. Ali et~al.,
\newblock Phys. Rev. {\bf D61} (2000) 074024, hep-ph/9910221.

\bibitem{Eichten:1990zv}
E. Eichten and B. Hill,
\newblock Phys. Lett. {\bf B234} (1990) 511.

\bibitem{Grinstein:1990mj}
B. Grinstein,
\newblock Nucl. Phys. {\bf B339} (1990) 253.

\bibitem{Georgi:1990um}
H. Georgi,
\newblock Phys. Lett. {\bf B240} (1990) 447.

\bibitem{Neubert:1994mb}
M. Neubert,
\newblock Phys. Rept. {\bf 245} (1994) 259, hep-ph/9306320.

\bibitem{Dugan:1991de}
M.J. Dugan and B. Grinstein,
\newblock Phys. Lett. {\bf B255} (1991) 583.

\bibitem{Lepage:1980fj}
G.P. Lepage and S.J. Brodsky,
\newblock Phys. Rev. {\bf D22} (1980) 2157.

\bibitem{Efremov:1980qk}
A.V. Efremov and A.V. Radyushkin,
\newblock Phys. Lett. {\bf B94} (1980) 245.

\bibitem{Chernyak:1990ag}
V.L. Chernyak and I.R. Zhitnitsky,
\newblock Nucl. Phys. {\bf B345} (1990) 137.

\bibitem{Khodjamirian:1997ub}
A. Khodjamirian et~al.,
\newblock Phys. Lett. {\bf B410} (1997) 275, hep-ph/9706303.

\bibitem{Bagan:1997bp}
E. Bagan, P. Ball and V.M. Braun,
\newblock Phys. Lett. {\bf B417} (1998) 154, hep-ph/9709243.

\bibitem{Szczepaniak:1990dt}
A. Szczepaniak, E.M. Henley and S.J. Brodsky,
\newblock Phys. Lett. {\bf B243} (1990) 287.

\bibitem{Burdman:1992hg}
G. Burdman and J.F. Donoghue,
\newblock Phys. Lett. {\bf B270} (1991) 55.

\bibitem{Akhoury:1994uw}
R. Akhoury, G. Sterman and Y.P. Yao,
\newblock Phys. Rev. {\bf D50} (1994) 358.

\bibitem{Dahm:1995ne}
M.~Dahm, R.~Jakob and P.~Kroll,
\newblock Z. Phys. {\bf C68} (1995) 595, hep-ph/9503418.

\bibitem{Braun:1990iv}
V.M. Braun and I.E. Filyanov,
\newblock Z. Phys. {\bf C48} (1990) 239.

\bibitem{Ball:1998sk}
P. Ball et~al.,
\newblock Nucl. Phys. {\bf B529} (1998) 323, hep-ph/9802299.

\bibitem{Feldmann:1999uf}
T. Feldmann,
\newblock Int. J. Mod. Phys. {\bf A15} (2000) 159, hep-ph/9907491.

\bibitem{Korchemsky:1999qb}
G.P. Korchemsky, D. Pirjol and T.M. Yan,
\newblock Phys. Rev. {\bf D61} (2000) 114510, hep-ph/9911427.

\bibitem{Ball:1998kk}
P. Ball and V.M. Braun,
\newblock Phys. Rev. {\bf D58} (1998) 094016, hep-ph/9805422.

\bibitem{Grozin:1997pq}
A.G. Grozin and M. Neubert,
\newblock Phys. Rev. {\bf D55} (1997) 272, hep-ph/9607366.

\bibitem{Gronberg:1998fj}
CLEO Collaboration, J. Gronberg et~al.,
\newblock Phys. Rev. {\bf D57} (1998) 33, hep-ex/9707031.

\bibitem{Kroll:1996jx}
P. Kroll and M. Raulfs,
\newblock Phys. Lett. {\bf B387} (1996) 848, hep-ph/9605264.

\bibitem{Ball:1998tj}
P. Ball,
\newblock JHEP {\bf 09} (1998) 005, hep-ph/9802394.

\bibitem{Burdman:1998mk}
G. Burdman,
\newblock Phys. Rev. {\bf D57} (1998) 4254, hep-ph/9710550.

\bibitem{Ali:1991is}
A. Ali, T. Mannel and T. Morozumi,
\newblock Phys. Lett. {\bf B273} (1991) 505.

\bibitem{Wandzura:1977qf}
S. Wandzura and F. Wilczek,
\newblock Phys. Lett. {\bf B72} (1977) 195.


\end{thebibliography}
\end{document}